\documentclass[journal,onecolumn,12pt,draftclsnofoot]{IEEEtran}

\usepackage{graphicx}
\usepackage{subfigure}
\usepackage{amstext}
\usepackage{amsthm,amsmath,amssymb}
\usepackage{bm} 
\usepackage{amsfonts}
\usepackage{microtype}

\usepackage{mathptmx}
\usepackage{amsthm,amsmath,amssymb,amsfonts}
\usepackage{bm} 
\usepackage{subeqnarray}
\usepackage{cases,cite}
\usepackage[utf8]{inputenc}
\usepackage{float}
\usepackage{graphicx} 
\usepackage{subfigure}
\usepackage{url}
\usepackage{multirow}
\usepackage{microtype}
\usepackage{booktabs}

\newtheorem{definition}{Definition}

\newtheorem{lemma}{Lemma}

\IEEEoverridecommandlockouts
\begin{document}

\title{Security of Distributed Machine Learning: A Game-Theoretic Approach to Design Secure DSVM}

\author{Rui~Zhang, Quanyan~Zhu

\thanks{Taken partially from the dissertation submitted to the Faculty of the New York University Tandon School of Engineering in partial fulfillment of the requirements for the degree Doctor of Philosophy, January 2020 \cite{zhang2020strategic}.

R. Zhang and Q. Zhu are with the Department of Electrical and Computer Engineering, New York University, Brooklyn, NY, 11201
E-mail:\{rz885,qz494\}@nyu.edu. }}

\maketitle

\begin{abstract}
Distributed machine learning algorithms play a significant role in processing massive data sets over large networks. However, the increasing reliance on machine learning on information and communication technologies (ICTs) makes it inherently vulnerable to cyber threats. This work aims to develop secure distributed algorithms to protect the learning from data poisoning and network attacks.  We establish a game-theoretic framework to capture the conflicting goals of a learner who uses distributed support vector machines (SVMs) and an attacker who is capable of modifying training data and labels. We develop a fully distributed and iterative algorithm to capture real-time reactions of the learner at each node to adversarial behaviors. The numerical results show that distributed SVM is prone to fail in different types of attacks, and their impact has a strong dependence on the network structure and attack capabilities.
\end{abstract}

\section{Introduction}
\label{Introduction}
Recently, parallel and distributed machine learning (ML) algorithms have been developed to scale up computations in large datasets and networked systems \cite{bekkerman2011scaling}, such as distributed spam filtering \cite{metzger2003multiagent} and distributed traffic control \cite{camponogara2003distributed}. However, they are inherently vulnerable to adversaries who can exploit them. For instance, nodes of distributed spam filter system can fail to detect spam email messages after an attacker modifies the training data \cite{nelson2009misleading}, or disrupts the network services using denial-of-service attacks \cite{wood2002denial}.

ML algorithms are often open-source tools and security is usually not the primary concerns of their designers. It is undemanding for an adversary to acquire the complete information of the algorithm, and exploits its vulnerabilities. Also, the growing reliance of ML algorithms on off-the-shelf information and communication technologies (ICTs) such as cloud computing and the wireless networks \cite{drevin2007value} has made it even easier for adversaries to exploit the existing known vulnerabilities of ICTs to achieve their goals. Security becomes a more critical issue in the paradigm of distributed ML, since the learning consists of a large number of nodes that communicate using ICTs, and its attack surface grows tremendously compared to its centralized counterpart.

Hence, it is imperative to design secure distributed ML algorithms against cyber threats. Current research endeavors have focused on two distinct directions. One is to develop robust algorithms despite uncertainties in the dataset \cite{globerson2006nightmare,dekel2010learning,maaten2013learning}. The second one is to improve detection and prevention techniques to defend against cyber threats, e.g., \cite{serpanos2001defense,levine2006survey,kohno2009secure}. These approaches have mainly focused on centralized ML and computer networks, separately. The investigation of security issues in the distributed ML over networks is lacking.

The challenges of developing secure distributed machine learning arise from the complexity of the adversarial behaviors and the network effects of the distributed system.  The attacker's strategies can be multi-stage. For instance, he can reach his target to modify training data by launching multiple successful network attacks. Furthermore, the impact of an attack can propagate over the network. Uncompromised nodes can be affected by the misinformation from compromised nodes, leading to a cascading effect.

Traditional detection and defense strategies for centralized ML and networked systems are not sufficient to protect the distributed ML systems from attacks. To bridge this gap, this work aims to develop a game-theoretic framework to model the interactions of an attacker and a defender to assess and design security strategies in distributed systems. Game theory has been used to address the security issues in centralized  ML, e.g., \cite{liu2009game,kantarciouglu2011classifier}, and those in computer networks \cite{michiardi2002game,lye2005game,huang2019differential} and cyber-physical networks \cite{chen2019dynamic,chen2019interdependent,chen2019optimal,nugraha2019subgame}. The proposed game-theoretic model captures the network structures of the distributed system and leads to fully distributed and iterative algorithms that can be implemented at each node as defense strategies. 

In particular, we use game models to study the impact on consensus-based distributed support vector machines (DSVMs) of an attacker who is capable of modifying training data and labels. {In \cite{zhang2018tnnls,zhang2015fusion,zhang2016student}, we have built a game-theoretic framework to investigate the impacts of data poisoning attacks to DSVMs, we have further proposed four defense methods and verified their effectiveness with numerical experiments in \cite{zhang2019jaif}. In \cite{zhang2017ciss}, we have proposed a game-theoretic framework to model the interactions between a DSVM learner and an attacker who can modify the training labels.}

{In this paper, we extend our previous works by studying a broad range of attack models and classify them into two types. One is ML attacks, which exploit the vulnerabilities of ML algorithm, and the other one is Net-attacks, which arise from the vulnerabilities of the communication network. We then build the game-theoretic minimax problem to capture the competition between a DSVM learner and an attacker who can modify training data and labels. With alternating direction method of multipliers (ADMM) \cite{boyd2011distributed,zhang2018ciss,zhang2019tsipn}, we develop a fully distributed iterative algorithms where each compromised node operates its own sub-max-problem for the attacker and sub-min-problem for the learner.} The game between a DSVM learner and an attacker can be viewed as a collection of small sub-games associated with compromised nodes. This unique structure enables the analysis of per-node-behaviors and the transmission of misleading information under the game framework.

{Numerical results on Spambase data set \cite{Spambase} are provided to illustrate the impact of different attack models by the attacker. We find that network plays a significant role in the security of DSVM, a balanced network with fewer nodes and higher degrees are less prone to attackers who can control the whole system. We also find that nodes with higher degrees are more vulnerable. The distributed ML systems are found to be prone to network attacks even though the attacker only adds small noise to information transmitted between neighboring nodes. A summary of notations in this paper is provided in the following table.}

\begin{table}[http]{
	\begin{tabular}{cc}
		\hline
		\multicolumn{2}{c}{Summary of Notations} 
		\\ \hline 
		\multicolumn{1}{c|}{$\mathcal{V}$, $v$, $\mathcal{B}_v$} & Set of Nodes, Node $v$, Set of Neighboring Nodes of Node $v$
		\\ \multicolumn{1}{c|}{$\mathbf{w}_v$, $b_v$, $\mathbf{r}_v$} & Decision Variables at Node $v$
		\\ \multicolumn{1}{c|}{$\mathbf{x}_{vn}$, $y_{vn}$} & $n$-th Data and Label at Node $v$
		\\ \multicolumn{1}{c|}{$\mathbf{X}_v$, $\mathbf{Y}_v$} & Data Matrix and Label Matrix at Node $v$
		\\ \multicolumn{1}{c|}{$\omega_{vu}$} & Consensus Variable between Node $v$ and Node $u$
		\\ \multicolumn{1}{c|}{$\theta_v$} & Indicator Vector of Flipped Labels at Node $v$
		\\ \multicolumn{1}{c|}{$\delta_{vn}$} & Vector of Data Poisoning on the $n$-th Data at Node $v$
		\\ \hline
	\end{tabular}}
\end{table}

\section{Preliminary}
\label{Preliminary}
\begin{figure}[]
\centering
\subfigure{
\includegraphics[width=0.4\textwidth]{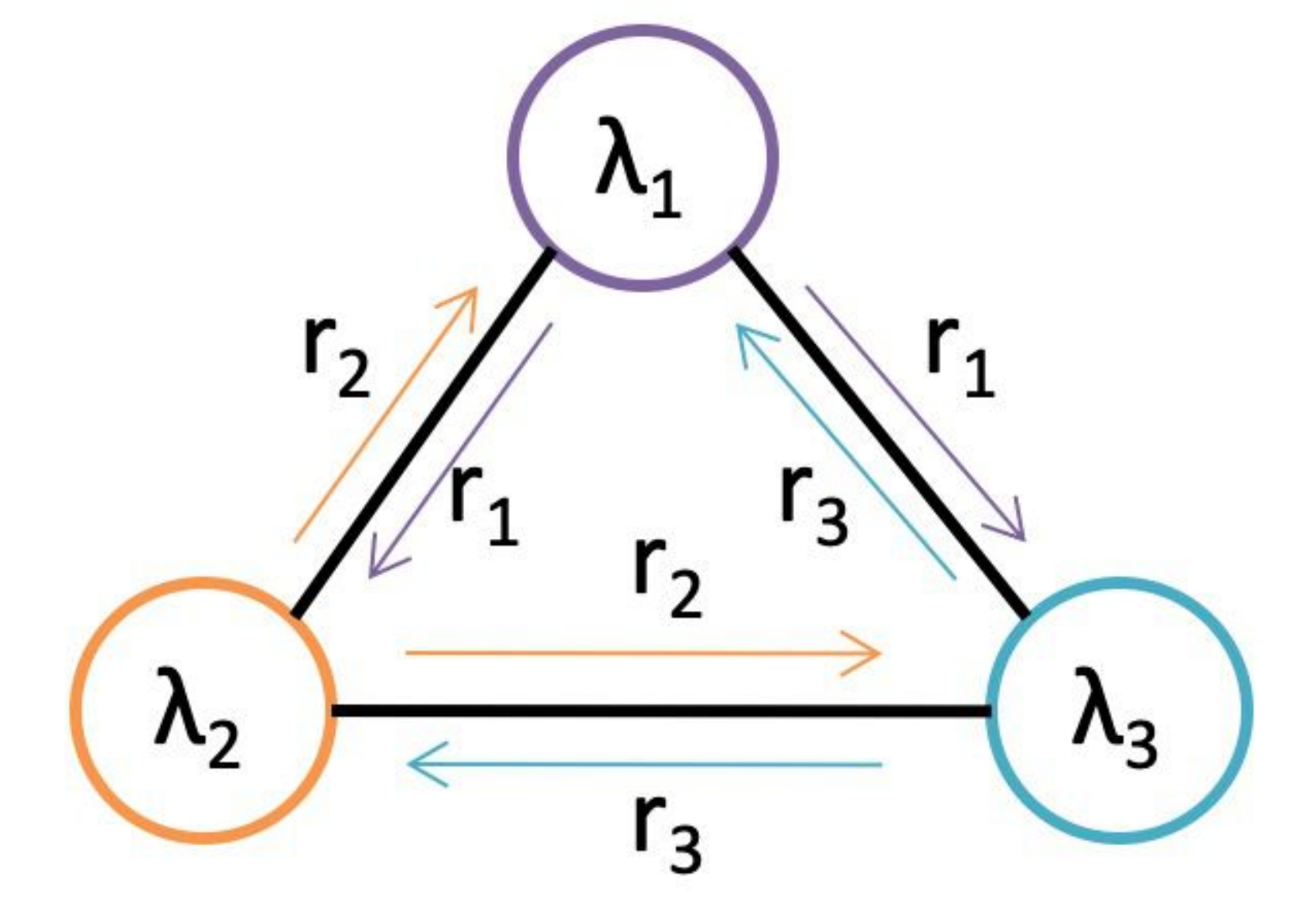}}
\subfigure{
\includegraphics[width=0.43\textwidth]{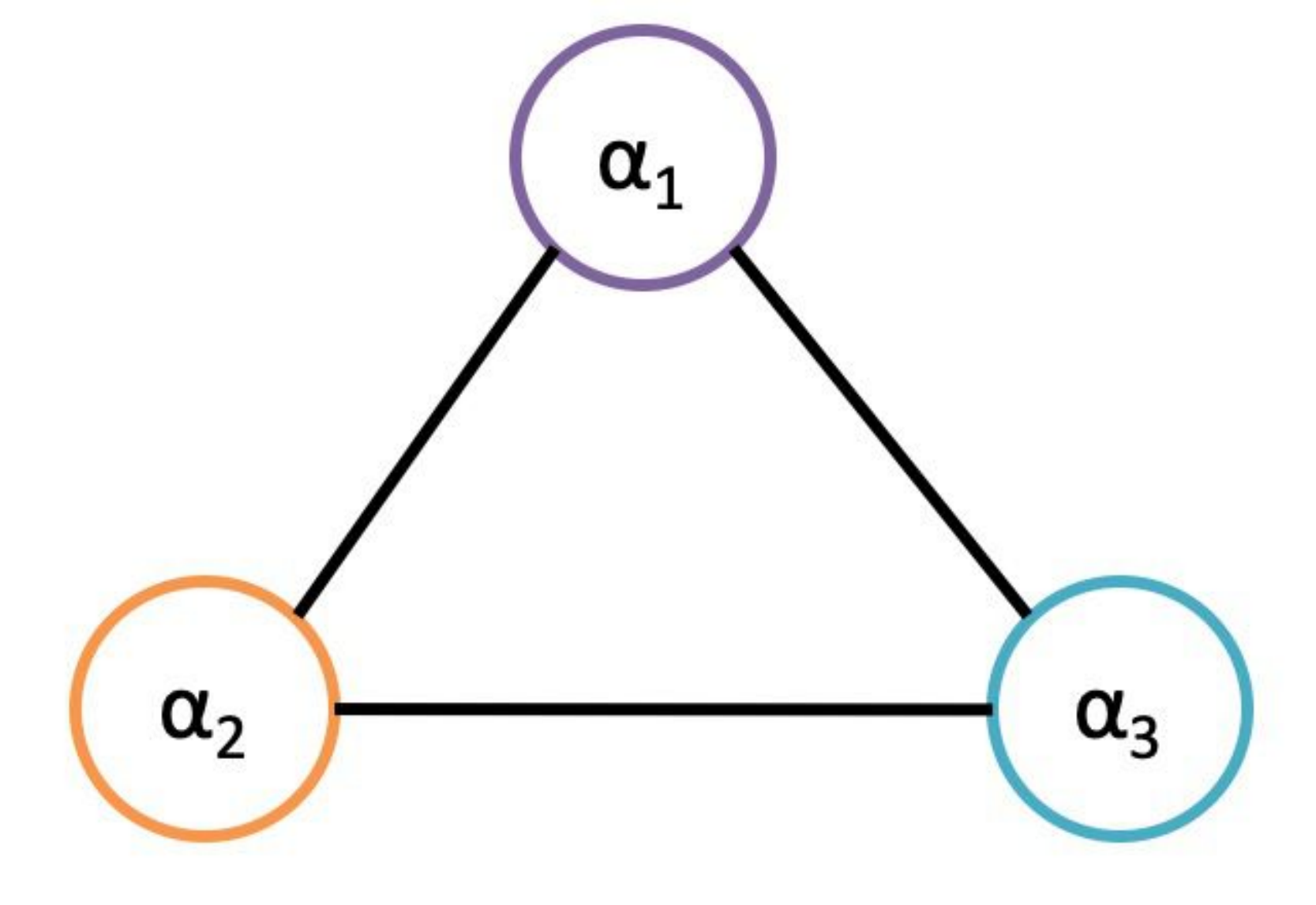}}
\caption{DSVM example \cite{zhang2017ciss}. At every iteration, each node $v$ first computes $\lambda_v$ and $\mathbf{r}_v$ by (\ref{eq:DSVMSoli1}) and (\ref{eq:DSVMSoli2}). Then, each node sends its $\mathbf{r}_v$ to its neighboring nodes. After that, each node computes $\alpha_v$ using (\ref{eq:DSVMSoli3}). Iterations continue until convergence. }
\label{fig:DSVM}
\end{figure}

Consider a distributed linear support vector machines learner in the network with $\mathcal{V}= \{1,..,V \}$ representing the set of nodes. Node $v\in\mathcal{V}$ only communicates with his neighboring nodes $\mathcal{B}_v \subseteq \mathcal{V}$. Note that without loss of generality, any two nodes in this network are connected by a path, i.e., there is no isolated node in this network. At every node $v$, a labelled training set $\mathcal{D}_v =\left\lbrace (\mathbf{x}_{vn},y_{vn}) \right\rbrace_{n = 1}^{N_v}$  of size $N_v $ is available. The goal of the learner is to find a maximum-margin linear discriminant function $g_v(\mathbf{x}) = \mathbf{x}^T\mathbf{w}_v^* + b_v*$ at every node $v\in\mathcal{V}$ based on its local training set $\mathcal{D}_v$. Consensus constraints $\mathbf{w}_1^* = \mathbf{w}_2^* = ... = \mathbf{w}_V^*, b_1^* = b_2^* = ... = b_V^* $ are used to force all local decision variables $\{\mathbf{w}_v^*,b_v^*  \}$ to agree across neighboring nodes. This approach enables each node to classify any new input $\mathbf{x}$ to one of the two classes $\{+1,-1 \}$ without communicating $\mathcal{D}_v$ to other nodes $v' \neq v$. The discriminant function $g_v(\mathbf{x})$ can be obtained by solving the following optimization problem: 
\begin{equation}
\label{eq:DSVM}
\begin{array}{c}
\min\limits_{ \{  \mathbf{w}_v,b_v \}  } \frac{1}{2} \sum\limits_{v\in\mathcal{V}} \parallel \mathbf{w}_v \parallel^2 + VC_l  \sum\limits_{v\in\mathcal{V}} \sum\limits_{n = 1}^{N_v} \left[ 1 - y_{vn}(\mathbf{w}_v^T\mathbf{x}_{vn} + b_{v})   \right]_+ \\
\begin{array}{lcr}
{\text{s.t.}}&{\mathbf{w}_v = \mathbf{w}_u,b_v = b_u,}&{\forall v\in \mathcal{V},u\in\mathcal{B}_v.}
\end{array}
\end{array}
\end{equation}
In the above problem, the term $\left[ 1 - y_{vn}(\mathbf{w}_v^T\mathbf{x}_{vn} + b_{v})   \right]_+  := \max\left[ 0,  1 - y_{vn}(\mathbf{w}_v^T\mathbf{x}_{vn} + b_{v}) \right] $ is the hinge loss function, $C_l $ is a tunable positive scalar for the learner. To solve Problem (\ref{eq:DSVM}), we first define $\mathbf{r}_v:=[{\bf{w}}_v^T,b_v]^T$, the augmented matrix $\mathbf{X}_v:=[(\mathbf{x}_{v1},...,\mathbf{x}_{vN_v})^T,\mathbf{1}_v]$, and the diagonal label matrix $\mathbf{Y}_v:=diag([y_{v1},...,y_{vN_v}])$. With these definitions, it follows readily that ${\bf{w}}_v=\widehat{\mathbf{I}}_{p \times(p+1)}\mathbf{r}_v$, $\widehat{\mathbf{I}}_{p \times(p+1)}=[{\mathbf{I}}_{p\times p},\mathbf{0}_{p\times  1}]$ is a $p \times (p+1)$ matrix with its first $p$ columns being an identity matrix, and its $(p+1)$ column being a zero vector. Thus, Problem (\ref{eq:DSVM}) can be rewritten as
\begin{equation}
\label{eq:DSVMMatrix}
\begin{array}{l}
\min\limits_{ \{  \mathbf{r}_v,\omega_{vu} \}  } \frac{1}{2} \sum\limits_{v\in\mathcal{V}} \mathbf{r}_v^T \Pi_{p+1} \mathbf{r}_v  + VC_l  \sum\limits_{v\in\mathcal{V}}  \mathbf{1}_v^T\left\lbrace\mathbf{1}_v -\mathbf{Y}_v\mathbf{X}_v\mathbf{r}_v \right\rbrace_{+}  \\
\begin{array}{lcc}
{\text{s.t.}}&{\begin{array}{c}
\mathbf{r}_v = \omega_{vu},\omega_{vn}=\mathbf{r}_u,
\end{array}}&{\begin{array}{c}
v\in \mathcal{V},u\in\mathcal{B}_v,
\end{array} }
\end{array}
\end{array}
\end{equation}
where the consensus variable $\omega_{vu}$ is used to decompose the decision variable $\mathbf{r}_v$ to its neighbors $\mathbf{r}_u$. Note that $\Pi_{p+1} = \widehat{\mathbf{I}}_{p \times(p+1)}^T\widehat{\mathbf{I}}_{p \times(p+1)}$ is a $(p+1)\times(p+1)$ identity matrix with its $(p+1,p+1)$-st entry being $0$. The term $\left\lbrace \mathbf{1}_v -\mathbf{Y}_v\mathbf{X}_v\mathbf{r}_v \right\rbrace_+  := \max\left[ \mathbf{0}_v,  \mathbf{1}_v -\mathbf{Y}_v\mathbf{X}_v\mathbf{r}_v\right]$, which returns a vector of size $N_v$. The algorithm of solving Problem (\ref{eq:DSVM}) can be shown as the following lemma from Proposition 1 in \cite{forero2010consensus}. 
\begin{lemma}
\label{lemma 1}
With arbitrary initialization $\mathbf{r}_v^{(0)},\lambda_v^{(0)}$ and $\alpha_v^{(0)}=\mathbf{0}_{(p+1)\times 1}$, the iterations per node are:
\begin{equation}
\label{eq:DSVMSoli1}
\begin{array}{*{20}{l}}
{{\lambda _v^{(t+1)}}}
{ \in \arg \mathop {\max }\limits_{{\bf{0}} \le {{\bf{\lambda }}_v} \le VC_l{{\bf{1}}_v}}  - \frac{1}{2}\lambda _v^T{{\bf{Y}}_v}{{\bf{X}}_v}{\bf{U}}_v^{ - 1}{\bf{X}}_v^T{{\bf{Y}}_v}{\lambda _v}}
{ + {{({{\bf{1}}_v} + {{\bf{Y}}_v}{{\bf{X}}_v}{\bf{U}}_v^{ - 1}{{\bf{f}}_v^{(t)}})}^T}{\lambda _v}},
\end{array}
\end{equation}
\begin{equation}
\label{eq:DSVMSoli2}
{{{\bf{r}}_v^{(t+1)}} = {\bf{U}}_v^{ - 1}\left( {{\bf{X}}_v^T{{\bf{Y}}_v}{\lambda _v^{(t+1)}} - {{\bf{f}}_v^{(t)}}} \right)},
\end{equation}
\begin{equation}
\label{eq:DSVMSoli3}
{{\alpha _v^{(t+1)}} = {\alpha _v^{(t)}} + \frac{\eta }{2}\sum\limits_{u \in {\mathcal{B}_v}} {\left[ {{{\bf{r}}_v^{(t+1)}} - {{\bf{r}}_u^{(t+1)}}} \right]} },
\end{equation}
where $\mathbf{U}_v=\Pi_{p+1}+2\eta\vert \mathcal{B}_v\vert\mathbf{I}_{p+1},\mathbf{f}_v^{(t)}=2\alpha_v^{(t)}-\eta\sum_{u\in \mathcal{B}_v}(\mathbf{r}_v^{(t)} + \mathbf{r}_u^{(t)})$.
\end{lemma}
Note that $\omega_{vu}$ has been solved directly and plugged into each equations. $\lambda_v$ and $\alpha_v$ are Lagrange multipliers. The ADMM-DSVM algorithm is illustrated in Figure \ref{fig:DSVM}. Note that at any given iteration $t$ of the algorithm, each node $v\in \mathcal{V}$ computes its own local discriminant function $g_v^{(t)} (\mathbf{x}) $ for any vector $\mathbf{x}$ as ${g_v^{(t)}({\bf{x}}) = [{{\bf{x}}^T},1]{{\bf{r}}_v^{(t)}}}$. Since we only need the decision variables $\mathbf{r}_v$ for the discriminant function, we use the $\mathbf{DSVM}_v$ as a short-hand notation to represent iterations (\ref{eq:DSVMSoli1})-(\ref{eq:DSVMSoli3}) at node $v$:
\begin{equation}
\label{eq:DSVMRe}
\mathbf{r}_v^{(t+1)} \in \mathbf{DSVM}_v\left( \mathbf{X}_v,\mathbf{Y}_v ,\mathbf{r}_v^{(t)},\{\mathbf{r}_u^{(t)}\}_{u\in\mathcal{B}_v} \right).
\end{equation}

The iterations stated in Lemma 1 are referred to as ADMM-DSVM. It is a fully decentralized network operation. Each node $v$ only shares decision variables $\mathbf{r}_v$ to his neighboring nodes $u\in\mathcal{B}_v$. Other DSVM approaches include distributed chunking SVMs \cite{navia2006distributed} and distributed parallel SVMs \cite{lu2008distributed}, where support vectors (SVs) are exchanged between each nodes, and distributed semiparametric SVMs, where the centroids of local SVMs are exchanged among neighboring nodes \cite{navia2006distributed}. In comparison, ADMM-DSVM has no restrictions on the network topology or the type of data, and thus, we use it as an example to illustrate the impact of the attacker on distributed learning algorithms. 
\section{Attack Models and Related Works}
In this section, we summarize and analyze possible attack models from the attacker. We start by identifying the attacker's goal and his knowledge. Based on the three critical characteristics of information, the attacker's goal can be captured as damageing the confidentiality, integrity and availability of the systems \cite{mccumber1991information}. Damaging confidentiality indicates that the attacker intends to acquire private information. Damaging integrity means that the data or the models used by the learner are modified or replaced by the attacker, which can not represent real information anymore. Damaging availability indicates that the attacker, which is an unauthorized user, uses the information or services provided by the learner. In this paper, we assume that the attacker intends to damage the integrity of the distributed ML systems by modifying either the data or the models of the learner. 

The impacts of the attacker are affected by the attacker's knowledge of the learner's systems. For example, an attacker may only know the data used by the learner, but he does not know the algorithm the learner use; or he may only know some nodes of the network. To fully capture the damages caused by the attacker, in this paper, we assume that the attacker has a complete knowledge of the learner, i.e., the attacker knows the learner's data and algorithm and the network topology.  

The attack models on distributed ML learner can be summarized into two main categories. One is the machine learning type attacks (ML-attacks) \cite{barreno2010security}, the other one is the network type attacks (Net-attacks) \cite{chi2001network}. In the ML-attacks, the attacker can exploit machine learning systems which produces classification or prediction errors. In the Net-attacks, an adversary attacks a networked system to compromise the security of this system by actions, which leads to the leak of private information or the failure of operations in this network. In this paper, we further divide ML-attacks into two sub-categories, training attacks and testing attacks. Note that the attack models described here are generally applicable to different machine learning algorithms. The focus of this work is to investigate the impact of these attack models on DSVM, which provides fundamental insights on the inherent vulnerability of distributed machine learning.
\subsection{Training Attacks}
In the training attacks, an adversary attacks the learner at the time when the learner solves Problem (\ref{eq:DSVM}). In these attacks, communications in the network may lead to unanticipated results as misleading information from compromised nodes can be spread to and then used by uncompromised nodes. One challenge of analyzing training attacks is that the consequences of attacker's actions may not be directly visible. For example, assuming that the attacker modifies some training data $\mathbf{x}_{vn}$ in  node $v$, the learner may not be able to find out which data has been modified, and furthermore, in distributed settings, the learner may not even be able to detect which nodes are under attack. We further divide training attacks into three categories based on the scope of modifications made by the attacker. 
\subsubsection{Training Labels Attacks}
In this category, the attacker can modify the training labels $\{ y_{vn} \}_{n = 1,...,N_v}$, where $v\in\mathcal{V}_a$. After training data with flipped labels, the discriminant functions will be prone to give wrong labels to the testing data. In early works \cite{biggio2011support,xiao2012adversarial},  centralized SVMs under training label attacks have been studied, and robust SVMs have been brought up to reduce the effects of such attacks. In this work, we further extend such attack models to distributed SVM algorithms, and we use game theory to model the interactions between the DSVM learner and the attacker. We verify the effects of the attacker with numerical experiments. 
\subsubsection{Training Data Attacks}
In this category, the attacker modifies the training data $\{ \mathbf{x}_{vn}\}_{n=1,...,N_v}$ on compromised nodes $v\in\mathcal{V}_a$. Since the training and testing data are assumed to be generated from the same distribution, the discriminant function found with training data on distribution $\mathcal{X}$ can be used to classify testing data from the same distribution $\mathcal{X}$. However, after modifying training data $\mathbf{x}_{vn}$ into $\widehat{\mathbf{x}}_{vn} $, which belongs to a different distribution $\widehat{\mathcal{X}}$, the discriminant function with training such crafted data is suitable to classify data of distribution $\widehat{\mathcal{X}}$. Thus, the testing data $\mathbf{x}\in\mathcal{X}$ are prone to be misclassified with this discriminant function.

The attacker can delete or craft several features \cite{dekel2010learning,maaten2013learning}, change the training data of one class \cite{dalvi2004adversarial}, add noise to training data \cite{biggio2011bagging}, change the distributions of training data \cite{liu2009game}, inject crafted data \cite{biggio2012poisoning}, and so on. However, these works aim at centralized machine learning algorithms. In distributed algorithms, the information transmissions between neighboring nodes can make uncompromised nodes to misclassify testing data after training with information from compromised nodes. In \cite{zhang2015fusion}, an attacker aims at reducing a DSVM learner's classification accuracy by changing training data $\mathbf{x}_{vn}$ in node $v\in\mathcal{V}_a$ into $\widehat{\mathbf{x}}_{vn}:= \mathbf{x}_{vn} -\delta_{vn} $.  This work shows that the performances of DSVM under adversarial environments are highly dependent on network topologies. 
\subsubsection{Training Models Attacks}
The DSVM learner aims to find the discriminant function with the lowest risks by solving Problem (\ref{eq:DSVM}) with local training sets $\mathcal{D}_v$. However, the attacker may change Problem (\ref{eq:DSVM}) into a different problem or he can modify parameters in Problem (\ref{eq:DSVM}). For example, when the attacker changes $C_l$ in Problem (\ref{eq:DSVM}) into $0$, the learner can only find $\mathbf{w}_v = \mathbf{0}$, which does not depend on the distribution of training data, and thus, the learner will misclassify input testing data in the same distribution. 

With training attacks, the DSVM leaner will find wrong decision variables $\widehat{\mathbf{w}}_v^*$ and $\widehat{b}_v^*$ in compromised node $v\in\mathcal{V}_a$. However, the consensus constraints $\mathbf{w}_1^* = \mathbf{w}_2^* = ... = \mathbf{w}_V^*, b_1^* = b_2^* = ... = b_V^* $ force all the decision variables to agree on each other. Hence uncompromised nodes with correct decision variables will be affected by misleading decision variables from compromised nodes. As a result, the training process in the network can be damaged even the attacker only attacks a small number of nodes. 
\subsection{Testing Attacks}
In testing attacks, the attacker attacks at the time when the DSVM learner labels new input $\mathbf{x}$ into $+1$ or $-1$ with $\mathbf{w}_v^*$ and $b_v^*$ from the solved Problem (\ref{eq:DSVM}). The attacker can conduct three different operations in testing attacks. 

Firstly, the attacker can directly change the given label $y$ of the testing data $\mathbf{x}$ into $-y$, and thus a wrong label is generated with this operation. Secondly, the attacker can replace the testing data $\mathbf{x}$ with crafted $\widehat{\mathbf{x}} $, or he can modify that into $\widehat{\mathbf{x}} = \mathbf{x} - \delta$ \cite{biggio2013evasion}. In such cases, the learner gives the label of $\widehat{\mathbf{x}} $ rather than the label of $\mathbf{x}$, which leads to misclassification. Thirdly, the attacker can modify or replace $\mathbf{w}_v^*$ and $b_v^*$ into $\widehat{\mathbf{w}}_v^*$ and $\widehat{b}_v^*$, and thus the compromised discriminant function $\widehat{g}_v(\mathbf{x}) = \mathbf{x}^T\widehat{\mathbf{w}}_v^* + \widehat{b}_v^*$ will yield wrong results. A simple example is that the attacker can set $\widehat{\mathbf{w}}_v^* = -\mathbf{w}_v^*$ and $\widehat{b}_v^* = -b_v^*$, thus $\widehat{g}_v(\mathbf{x}) = -g_v(\mathbf{x})$, which leads to contrary predictions. 

Testing attacks can cause disastrous consequences in centralized machine learning algorithms as there is only one discriminant function. However, Testing attacks are weak in distributed machine learning algorithms as uncompromised nodes can still give correct predictions, and compromised nodes can be easily detected as they have higher classification risks. 
\subsection{Network Attacks}
\begin{figure}[]
\centering
\subfigure[]{
\includegraphics[width=0.2\textwidth]{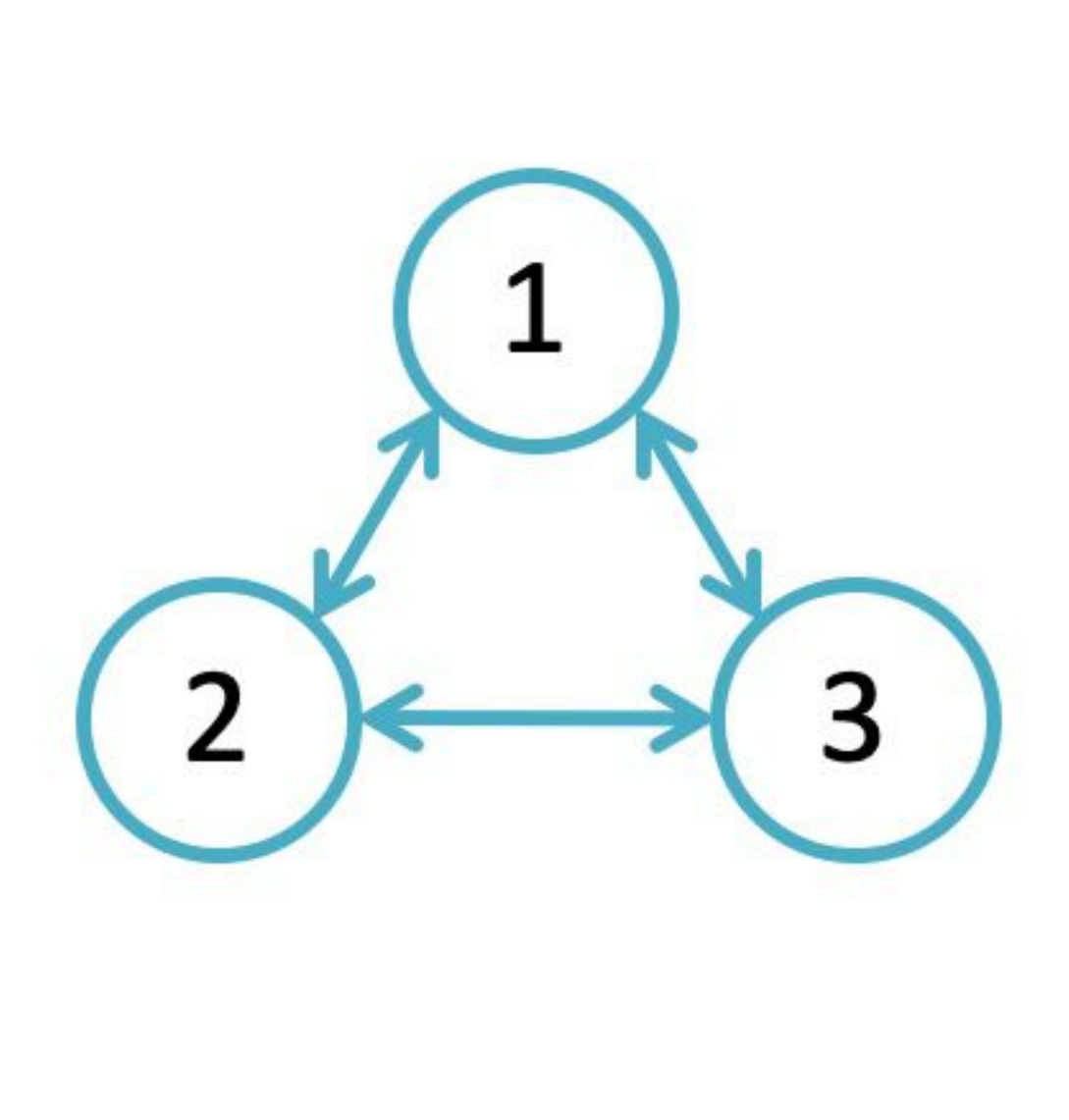}}
\subfigure[]{
\includegraphics[width=0.2\textwidth]{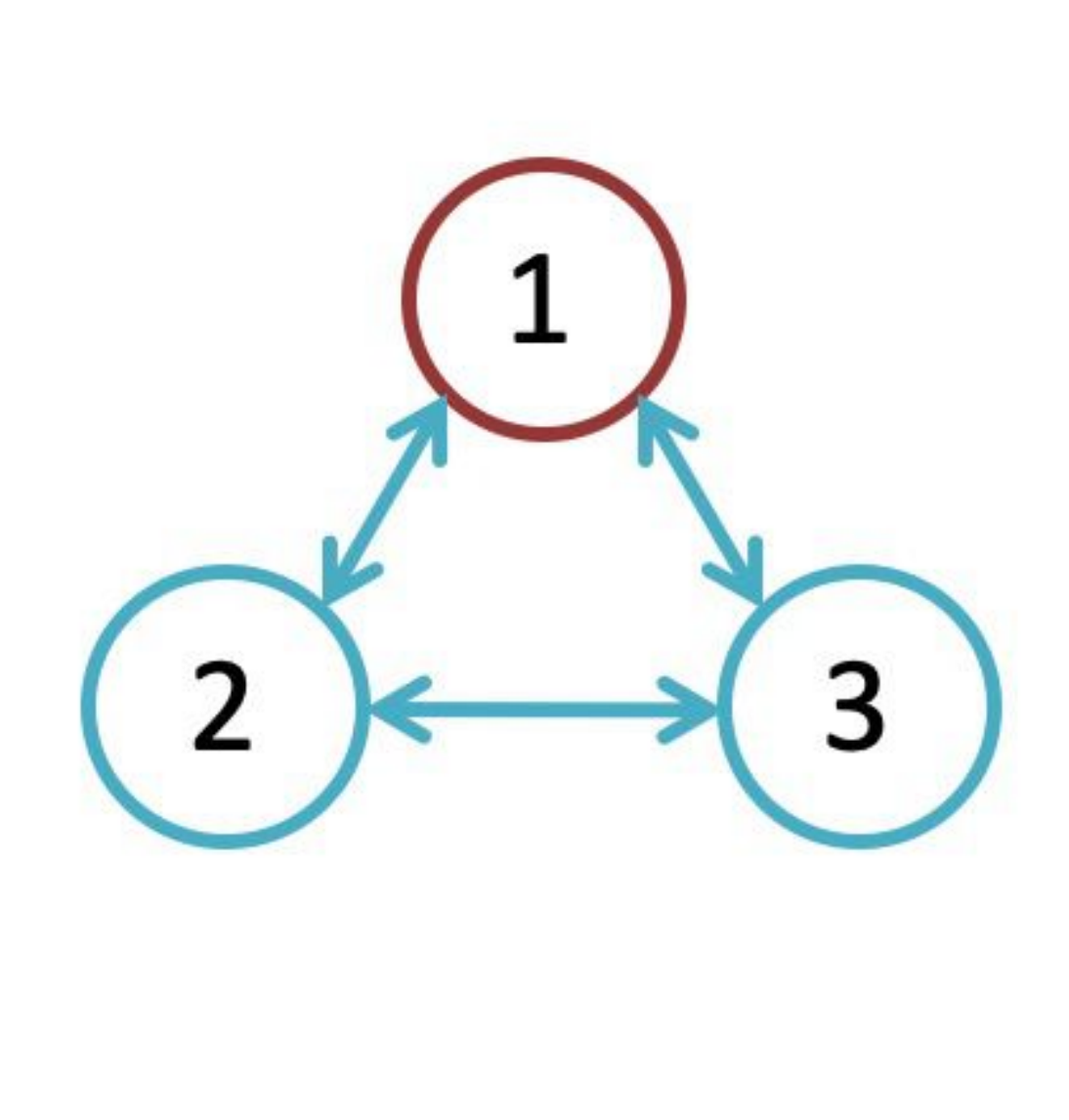}}
\subfigure[]{
\includegraphics[width=0.2\textwidth]{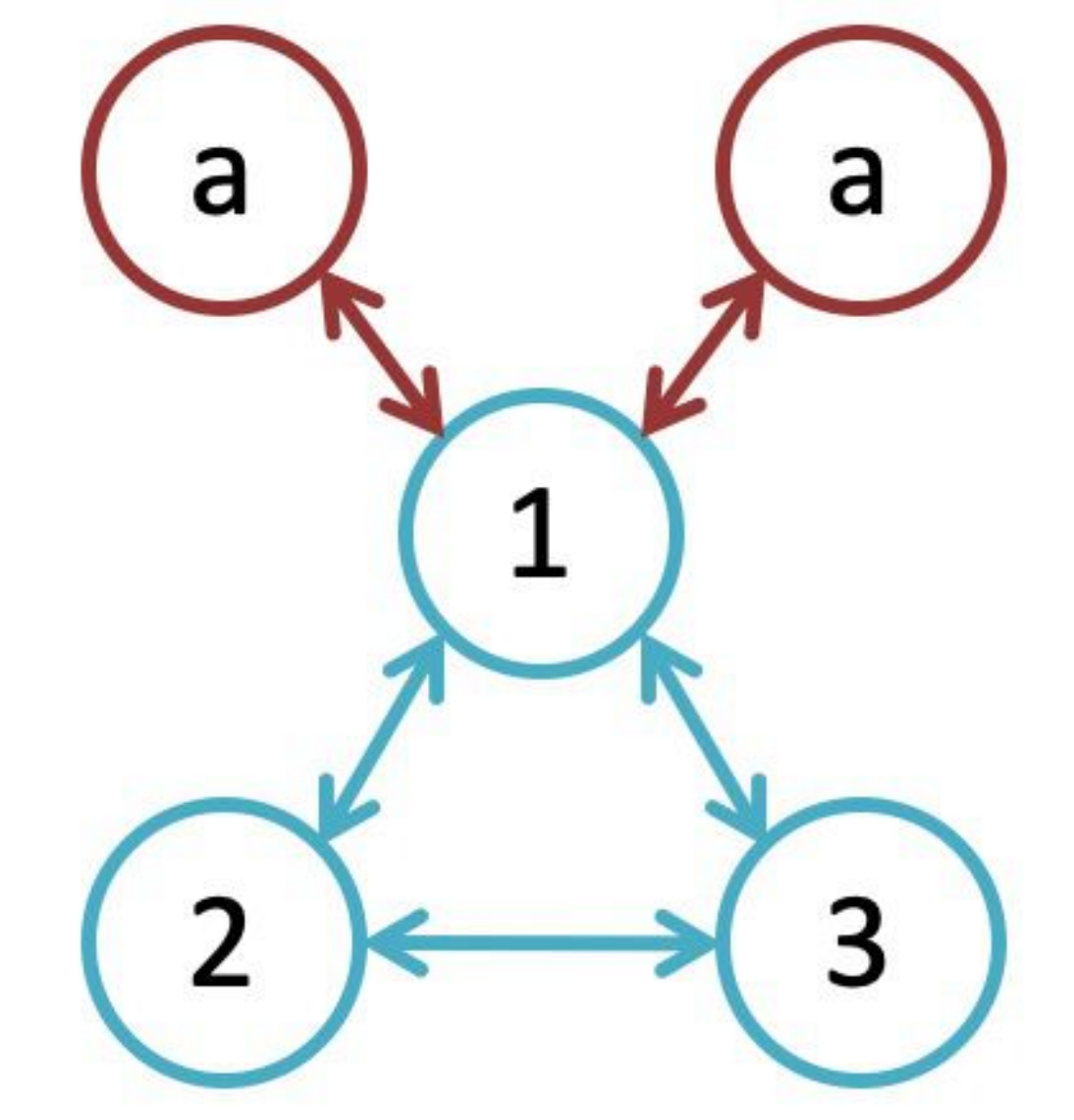}}
\subfigure[]{
\includegraphics[width=0.2\textwidth]{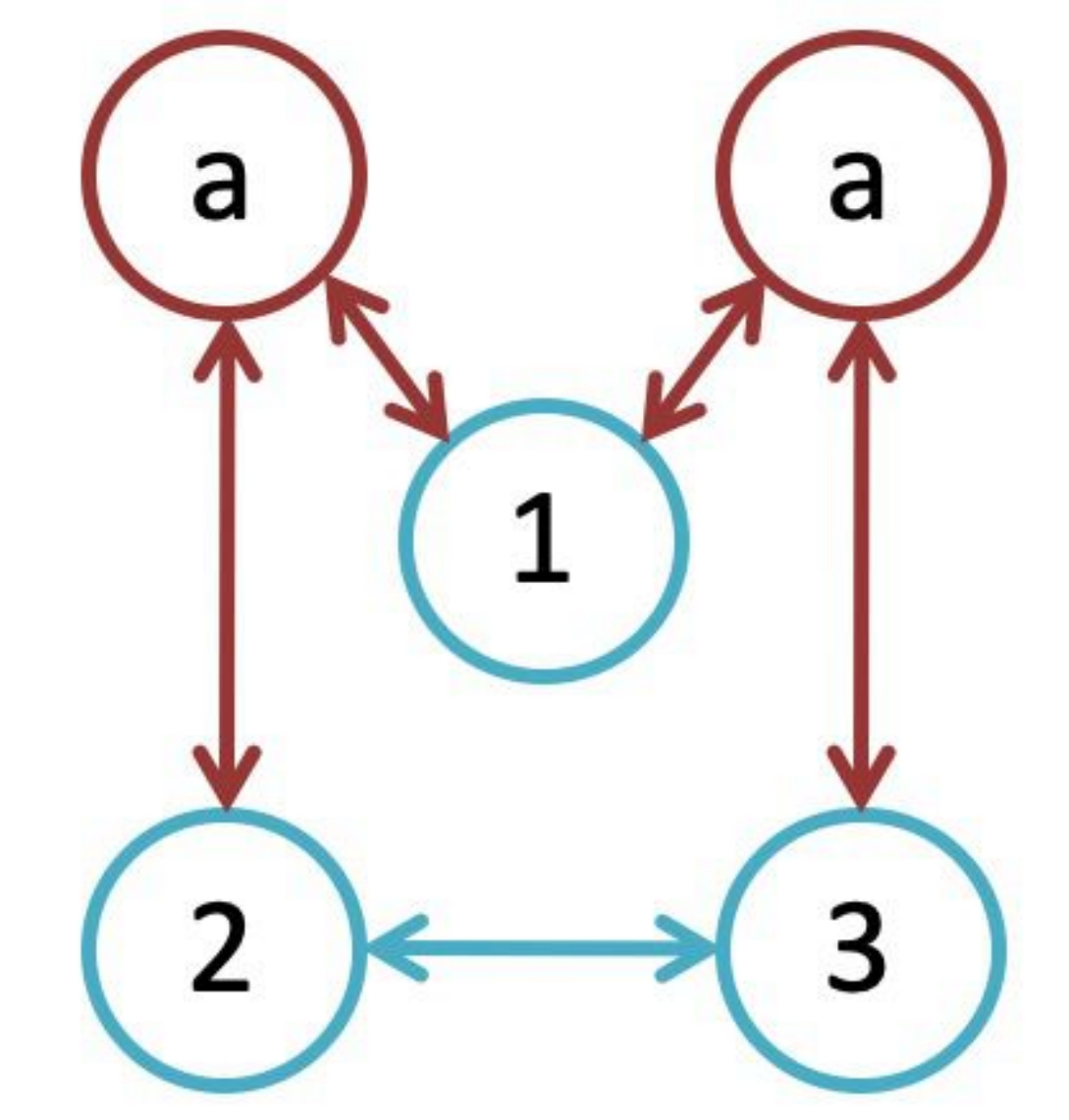}}
\caption{Network attacks examples. (a) No attack. The learner uses a fully connected network with $3$ nodes. (b) Node capture attacks. The attacker controls Node $1$ and make modifications on it. (c) Sybil attacks. Two red malicious nodes are generated by the attacker and pretend that they are in the training network. (d) MITM attacks. The attacker creates adversarial nodes (depicted in red) and pretend to be the other nodes on the connections.  }
\label{fig:CyberAttacks}
\end{figure}

Network attacks can pose a
significant threat with potentially severe consequences on networked systems \cite{chi2001network}. Since distributed machine learning algorithms have been developed to solve large-scale problems using networked systems \cite{forero2010consensus}, network attacks can also cause damages on distributed learning systems. Network attacks include node capture attacks, Sybil attacks, and Man-In-The-Middle attacks, which are illustrated in Figure \ref{fig:CyberAttacks}.

In node capture attacks (Figure \ref{fig:CyberAttacks}(b)), an attacker gains access into the network, and controls a set of nodes, then he can alter both software programming and hardware configuration, and influence the outcome of network protocols \cite{tague2008modeling}. When the attacker conducts node capture attacks on a DSVM learner, he can modify either the data in the compromised nodes, or the algorithms of the learner. Both of the modifications can lead to misclassifications in compromised nodes. Moreover, the attacker can also send misleading information through  network connections between neighboring nodes, thus even uncompromised nodes can be affected by the attacker. 

In Sybil attacks (Figure \ref{fig:CyberAttacks}(c)), the attacker can create an adversarial node in the middle of a connection, and talk to both nodes and pretends to be the other node \cite{newsome2004sybil}. If such attacks happen on the network of a DSVM learner, nodes in compromised connections will receive misleading information, and thus, the classifications in these nodes will be damaged.

In Man-in-the-Middle (MITM) attacks (Figure \ref{fig:CyberAttacks}(d)), the attacker can create an adversarial node in the middle of a connection, talk to both nodes and pretend to be the other node \cite{desmedt2011man}. If such attacks happen on the network of a DSVM learner, nodes in compromised connections will receive misleading information, and thus classifications in these nodes will be damaged. 

There are many other network attack models, such as botnet attacks \cite{zhang2011survey} and denial of service \cite{wood2002denial}, which makes it challenging to analyze and defend attacker's behaviors. Though distributed systems improve the efficiency of the learning algorithms, however, the systems becomes more vulnerable to network attacks. Thus, it is important to design secure distributed ML algorithms against potential adversaries.

With various types of attack models, a distributed machine learning learner can be vulnerable in a networked system. Thus, it is important to design secure distributed algorithms against potential adversaries. Since the learner aims to increase the classification accuracy, while the attacker seeks to reduce that accuracy, game theory becomes a useful tool to capture the conflicting interests between the learner and the attacker. The equilibrium of this game allows us to predict the outcomes of a learner under adversary environments. In the next section, we build a game-theoretic framework to capture the conflicts between a DSVM learner and an attacker who can modify training labels and data.
\section{A Game-Theoretic Modeling of Attacks}
In this section, we use training attacks as an example to illustrate the game-theoretic modeling of the interactions between a DSVM learner and an attacker. We mainly focus on modelling training labels attacks, and the training data attacks. 
\subsection{Training Labels Attacks}
In training labels attacks, the attacker controls a set of nodes $\mathcal{V}_a$, and aims at breaking the training process by flipping training labels $y_{vn}$ to $\widehat{y}_{vn} = :-y_{vn}$. To model the flipped labels at each node, we first define the matrix of expanded training data $\widehat{\mathbf{X}}_v : = [ [\mathbf{x}_{v1},...,\mathbf{x}_{vN_v},\mathbf{x}_{v1},...,\mathbf{x}_{vN_v}]^T, \widehat{\mathbf{1}}_v] $ and the diagonal matrix of expanded labels $\widehat{\mathbf{Y}}_v:=diag([y_{v1},...,y_{vN_v},-y_{v1},...,-y_{vN_v}])$. Note that the first $N_v$ data and the second $N_v$ data are the same in the matrix of expanded training data. We further introduce corresponding indicator vector $\mathbf{\theta}_v := [\theta_{v1},...,\theta_{v(2N_v)}]^T$, where $\theta_{vn} \in \{ 0, 1\}$, and $\theta_{vn}+\theta_{v(n+N_v)} = 1$, for $n = 1,...,N_v$ \cite{xiao2012adversarial}. $\theta_v$ indicates whether the label has been flipped, for example, if $\theta_{vn} = 0$ for $n = 1,...,N_v$, i.e., $\theta_{v(n+N_v)} = 1$, then the label of data $\mathbf{x}_{vn}$ has been flipped. Note that $\mathbf{\theta}_v = [\mathbf{1}_v^T,\mathbf{0}_v^T]^T$ indicates that there is no flipped label. 

Since the learner aims to minimize the classification errors by minimizing the objective function in Problem (\ref{eq:DSVMMatrix}), the attacker's intention to maximize classification errors can be captured as maximizing that objective function. As a result, the interactions of the DSVM learner and the attacker can be captured as a nonzero-sum game. Furthermore, since they have same objective functions with opposite intentions,  the nonzero-sum game can be reformulated into a zero-sum game, which takes the minimax or max-min form \cite{zhang2017ciss}:  
\begin{equation}
\label{eq:MinMaxMatrix}
\begin{array}{l}
\min\limits_{ \{  \mathbf{r}_v,\omega_{vu} \}  } \max\limits_{ \{  \theta_{v} \}_{v\in\mathcal{V}_a}  } K\left( {\left\{ {{{\bf{r}}_v,\omega_{vu}}} \right\},\left\{ {{\theta _{v}}} \right\}} \right):= \frac{1}{2} \sum\limits_{v\in\mathcal{V}} \mathbf{r}_v^T \Pi_{p+1} \mathbf{r}_v \\ 
\ \ \ \ \ \ \ \ \ \ \ \ \ \ + VC_l  \sum\limits_{v\in\mathcal{V}} \theta_{v}^T  \left\lbrace\widehat{\mathbf{1}}_v -\widehat{\mathbf{Y}}_v\widehat{\mathbf{X}}_v\mathbf{r}_v \right\rbrace_{+} - V_a C_a\sum\limits_{v\in\mathcal{V}_a}\theta_v^T [ \mathbf{0}_{v}^T,\mathbf{1}_{v}^T ]^T \\
\begin{array}{lccr}
{\text{s.t.}}&{\begin{array}{c}
\mathbf{r}_v = \omega_{vu},\omega_{vn}=\mathbf{r}_u,\\ \theta_v = [\mathbf{1}_v^T,\mathbf{0}_v^T]^T;\\
\mathbf{q}_{v}^T\mathbf{\theta}_{v} \leq Q_v,\\
\left[\mathbf{I}_{N_v},\mathbf{I}_{N_v}\right]\mathbf{\theta}_{v} = \mathbf{1}_{v},\\
\theta_{v} \subseteq \{ 0,1 \}_{2N_v}
\end{array}}&{\begin{array}{c}
v\in \mathcal{V},u\in\mathcal{B}_v;\\
v\in\mathcal{V}_l;\\
v\in \mathcal{V}_a;\\v\in\mathcal{V}_a;\\ v\in\mathcal{V}_a.
\end{array} }&{\begin{array}{c}
(\ref{eq:MinMaxMatrix}a)\\
(\ref{eq:MinMaxMatrix}b)\\
(\ref{eq:MinMaxMatrix}c)\\
(\ref{eq:MinMaxMatrix}d)\\
(\ref{eq:MinMaxMatrix}e)\\
\end{array}}
\end{array}
\end{array}
\end{equation}
In (\ref{eq:MinMaxMatrix}), $\widehat{\mathbf{1}}_v= [1,1,...,1]_{2N_v}^T$ denotes a vector of size $2N_v$, $\mathcal{V}_l$ denotes the set of uncompromised nodes. Note that the first two terms in the objective function and Constraints (\ref{eq:MinMaxMatrix}a) are related to the min-problem for the learner. When $\mathbf{\theta}_v = [\mathbf{1}_v^T,\mathbf{0}_v^T]^T$, i.e., there are no flipped labels, the min-problem is equivalent to Problem (\ref{eq:DSVMMatrix}). The last two terms in the objective function and Constraints (\ref{eq:MinMaxMatrix}c)-(\ref{eq:MinMaxMatrix}e) are related to the max-problem for the attacker. Note that the last term of the objection function represents the number of flipped labels in compromised nodes. By minimizing this term,  the attacker aims to create the largest impact by flipping the fewest labels. In Constraints (\ref{eq:MinMaxMatrix}c), $\mathbf{q}_v:=[0,..,0,q_{v(N_v+1)},...,q_{v(2N_v)}]^T$, where $\mathbf{q}_{v(N_v+n)}$ indicates the cost for flipping the labels of $\mathbf{x}_{vn}$ in node $v$. This constraint indicates that the capability of the attacker is limited to flip labels with a boundary $Q_v$ at a compromised node $v$. Constraints (\ref{eq:MinMaxMatrix}d) show that the labels in compromised nodes are either flipped or not flipped. 

The Minimax-Problem (\ref{eq:MinMaxMatrix}) captures the learner's intention to minimize the classification errors with attacker's intention of maximizing that errors by flipping labels. Problem (\ref{eq:MinMaxMatrix}) can be also written into a max-min form, which captures the attacker's intention to maximize the classification errors while the learner tries to minimize it. By minimax
theorem, the minimax form in (\ref{eq:MinMaxMatrix}) is equivalent to its max-min form. Thus, solving Problem (\ref{eq:MinMaxMatrix}) can be interpreted as finding the saddle-point equilibrium of the zero-sum game between the learner and the attacker.
\label{definitino1}
\begin{definition}
Let $\mathcal{S}_L $ and $ \mathcal{S}_A$ be the action sets for the DSVM learner and the attacker respectively. Then, the strategy pair $\left( {\left\{ {{{\bf{r}}_v^*},{\omega_{vu}^*}} \right\},\left\{ {{\theta_{v}^*}} \right\}} \right)$ is a saddle-point equilibrium solution of the zero-sum game defined by the triple ${G_z} := \left\langle {\left\{ {L,A} \right\},\left\{ {{\mathcal{S}_L},{\mathcal{S}_A}} \right\},K} \right\rangle $, if
$K\left( {\left\{ {{{\bf{r}}_v^*},{\omega_{vu}^*}} \right\},\left\{ {{\theta _{v}}} \right\}} \right) \leq   K\left( {\left\{ {{{\bf{r}}_v^*},{\omega_{vu}^*}} \right\},\left\{ {{\theta _{v}^*}} \right\}} \right) \leq K\left( {\left\{ {{{\bf{r}}_v},{\omega_{vu}}} \right\},\left\{ {{\theta _{v}^*}} \right\}} \right),\forall v \in \mathcal{V}_a $, 
where $K$ is the objective function in Problem (\ref{eq:MinMaxMatrix}).
\end{definition}
To solve Problem (\ref{eq:MinMaxMatrix}), we construct the best response dynamics for the max-problem and min-problem separately. The max-problem and the min-problem can be achieved by fixing $\{ \mathbf{r}_v^*,\omega_{vu}^*  \}$ and $\{ \theta_v^*  \}$, respectively. {With solving both problems in a distributed way, we achieve the fully distributed iterations of solving Problem (\ref{eq:MinMaxMatrix}) as}
\begin{equation}
\label{eq:MinMaxSoli1}
\begin{array}{l}
\theta_v^{(t+1)} \in \arg\max\limits_{   \theta_{v}   }  VC_l \theta_{v}^T  \lbrace\widehat{\mathbf{1}}_v -\widehat{\mathbf{Y}}_v\widehat{\mathbf{X}}_v\mathbf{r}_v^{(t)} \rbrace_{+}   - V_aC_a\theta_{v}^T [ \mathbf{0}_{v}^T,\mathbf{1}_{v}^T ]^T  \\
\begin{array}{lrr}
{\text{s.t.}}&{\begin{array}{c}
\mathbf{q}_{v}^T\mathbf{\theta}_{v} \leq Q_v,\\
\left[\mathbf{I}_{N_v},\mathbf{I}_{N_v}\right]\mathbf{\theta}_{v} = \mathbf{1}_{v},\\
\widehat{\mathbf{0}}_v \leq \theta_{v} \leq \widehat{\mathbf{1}}_v.
\end{array}}&{\begin{array}{c}
(\ref{eq:MinMaxSoli1}a)\\
(\ref{eq:MinMaxSoli1}b)\\
(\ref{eq:MinMaxSoli1}c)
\end{array}}
\end{array}
\end{array}
\end{equation} 
\begin{equation}
\label{eq:DSVMLRe}
\mathbf{r}_v^{(t+1)}\in\mathbf{DSVM}_{v,L}\left(\widehat{\mathbf{X}}_v,\widehat{\mathbf{Y}}_v,\mathbf{r}_v^{(t)},\{\mathbf{r}_u^{(t)}\}_{u\in\mathcal{B}_v} \big|\theta_v^{(t+1)}   \right)
\end{equation}
Problem (\ref{eq:MinMaxSoli1}) is a linear programming problem. Note that integer constraint (\ref{eq:MinMaxMatrix}e) has been further relaxed into (\ref{eq:MinMaxSoli1}c).
It captures the attacker's actions in compromised nodes $v\in\mathcal{V}_a$. Note that each node can achieve their own $\theta_v$ without transmitting information to other nodes. Problem (\ref{eq:DSVMLRe}) comes from the min-part of Problem (\ref{eq:MinMaxMatrix}), which can be solved using similar method in \cite{forero2010consensus} with ADMM \cite{boyd2011distributed}. Note that $\mathbf{DSVM}_{v,L}$ differs from $\mathbf{DSVM}_{v}$ only in the feasible set of the Lagrange multipliers $\lambda_v$, In $\mathbf{DSVM}_{v,L}$, ${{\bf{0}} \le {{\bf{\lambda }}_v} \le VC_l{\theta_v^{(t+1)}}}$, where $\theta_v^{(t+1)}$ indicates whether the label has been flipped, and it comes from the attacker's Problem (\ref{eq:MinMaxSoli1}). 

With (\ref{eq:MinMaxSoli1}) and (\ref{eq:DSVMLRe}), the algorithm of solving Problem (\ref{eq:MinMaxMatrix}) can be summarized: Each compromised node computes $\theta_v$ via (\ref{eq:MinMaxSoli1}), then compromised and uncompromised nodes compute $\mathbf{r}_v$ via (\ref{eq:DSVMRe}) and (\ref{eq:DSVMLRe}), respectively. The iterations will go until convergence. 
\subsection{Training Data Attacks}
In training data attacks, the attacker has the ability to modify the training data $\mathbf{x}_{vn}$ into $\widehat{\mathbf{x}}_{vn} := \mathbf{x}_{vn} - \delta_{vn}$ in compromised node $v\in\mathcal{V}_a$. Following a similar method in training label attacks, we can capture the interactions of the DSVM learner and the attacker as a zero-sum game which is shown as follows:
\begin{equation}
\label{eq:MinMaxData}
\begin{array}{l}
\begin{array}{*{20}{l}}
{\mathop {\min }\limits_{\left\{ {{{\bf{w}}_v},{b_v}} \right\}} \mathop {\max }\limits_{\left\{ {{\delta _{vn}}} \right\}} K\left( {\left\{ {{{\bf{w}}_v},{b_v}} \right\},\left\{ {{\delta _{vn}}} \right\}} \right) : = \frac{1}{2}\sum\limits_{v \in \mathcal{V}} {{{\left\| {{{\bf{w}}_v}} \right\|^2}}} }\\
{\begin{array}{*{20}{c}}
{}&{}
\end{array} + {V_l}{C_l}\sum\limits_{v \in \mathcal{V}_l} {\sum\limits_{n = 1}^{{N_v}} {{{\left[ {1 - {{\rm{y}}_{vn}}({\bf{w}}_v^T{{\bf{x}}_{vn}} + {b_v})} \right]}_ + }} } }\\
{\begin{array}{*{20}{c}}
{}&{}
\end{array} + {V_a}{C_l}\sum\limits_{v \in \mathcal{V}_a} {\sum\limits_{n = 1}^{{N_v}} {{{\left[ {1 - {{\rm{y}}_{vn}}({\bf{w}}_v^T({{\bf{x}}_{vn}} - {\delta _{vn}}) + {b_v})} \right]}_ + }} } }\\
{\begin{array}{*{20}{c}}
{}&{}
\end{array} - {V_aC_a}\sum\limits_{v \in \mathcal{V}_a} {\sum\limits_{n = 1}^{{N_v}} {{{\left\| {{\delta _{vn}}} \right\|}_0}} } }
\end{array}\\
{\rm{s}}{\rm{.t}}{\rm{.   }}\begin{array}{*{20}{c}}
{\begin{array}{*{20}{l}}
{{{\bf{w}}_v} = {{\bf{w}}_u},{b_v} = {b_u},}\\
{\sum_{n = 1}^{N_v}\parallel\delta_{vn}\parallel^2 \leq C_{v,\delta},}
\end{array}}&{\begin{array}{*{20}{l}}
{\forall v \in {\cal V},u \in {\mathcal{B}_v};}\\
{\forall v \in {{\cal V}_a}.}
\end{array}}&{\begin{array}{*{20}{l}}
{(\ref{eq:MinMaxData}a)}\\
{(\ref{eq:MinMaxData}b)}
\end{array}}
\end{array}
\end{array}
\end{equation}
Note that in the last term, $l_0$ norm ${\left\| x \right\|_0} := | \{i : {x_i} \neq 0\}|$ denotes the number of elements which are changed by the attacker, and deleting it captures the attacker's intentions to maximizing the classification errors with changing least number of elements. Constraint (\ref{eq:MinMaxData}b) indicates that the sum of modifiactions in node $v$ are bounded by $C_{v,\delta}$. Following a similar method in training labels attacks, we can construct the iterations of solving Problem (\ref{eq:MinMaxData}) as follows \cite{zhang2015fusion,zhang2016student,zhang2018tnnls}:
\begin{equation}
\label{eq:MinMaxDataMax}
\begin{array}{l}
{\delta _v^{(t+1)}} \in \arg \mathop {\max }\limits_{\left\{ {{\delta _v},{s_v}} \right\}} {V_a}{C_l}{\bf{r}}_v^{(t)T}\widehat{\mathbf{I}}_{p \times(p+1)}^T\delta _v   - {{\bf{1}}_v^T}{s_v}\\
{\rm{s}}.{\rm{t}}.{\rm{   }}\begin{array}{*{20}{cc}}
{\begin{array}{*{20}{c}}
{{V_aC_a}{\delta _v} \le {s_v},}\\
{{V_aC_a}{\delta _v} \ge  - {s_v},}\\
{\parallel \delta_v \parallel^2  \leq C_{v,\delta}.}
\end{array}}&{\begin{array}{*{20}{c}}
(\ref{eq:MinMaxDataMax}a)\\
(\ref{eq:MinMaxDataMax}b)\\
(\ref{eq:MinMaxDataMax}c)
\end{array}}
\end{array}
\end{array}
\end{equation}
\begin{equation}
\label{eq:DSVMDRe}
\mathbf{r}_v^{(t+1)}\in\mathbf{DSVM}_{v,D}\left({\mathbf{X}}_v,{\mathbf{Y}}_v,\mathbf{r}_v^{(t)},\{\mathbf{r}_u^{(t)}\}_{u\in\mathcal{B}_v} \big|\delta_v^{(t+1)}   \right)
\end{equation}
Note that here $\delta_{vn}$ has been summed into $\delta_v$, which captures the modifications in node $v$. Constraints (\ref{eq:MinMaxDataMax}ab) and the last term of the objective function is the relaxation of the $l_0$ norm. Note that comparing to $\mathbf{DSVM}_v$, $\mathbf{DSVM}_{v,D}$ has $\mathbf{f}_v^{(t+1)}=V_a C_l\delta_v^{(t+1)}+2\alpha_v^{(t)}-\eta\sum_{u\in \mathcal{U}_v}(\mathbf{r}_v^{(t)}+\mathbf{r}_u^{(t)})$, where $\delta_v^{(t+1)}$ comes from Problem (\ref{eq:MinMaxDataMax}).

In this section, we have modeled the conflicting interests of a DSVM learner and an attacker using a game-theoretic framework. The interactions of them can be captured as a zero-sum game where a minimax problem is formulated. The minimization part captures the learner's intentions to minimize classification errors, while the maximization part captures the attacker's intentions to maximize that errors with making less modifications, i.e., flipping labels and modifying data. By constructing the min-problem for the learner and max-problem for the attacker separately, the minimax problem can be solved with the best response dynamics. Furthermore, the min-problem and max-problem can be solved in a distributed way with $V_a$ Sub-Max-Problems (\ref{eq:MinMaxSoli1}) and (\ref{eq:MinMaxDataMax}), and $V$ Sub-Min-Problems (\ref{eq:DSVMLRe}) and (\ref{eq:DSVMDRe}). Combing the iterations of solving these problems, we have the fully distributed iterations of solving the Minimax-Problems (\ref{eq:MinMaxMatrix}) and (\ref{eq:MinMaxData}). The nature of this iterative operations provides real-time mechanisms for each node to reacts to its neighbors and the attacker. Since each node operates its own sub-max-problem and sub-min-problems, the game between a DSVM learner and an attacker now can be represented by $V_a$ sub-games in compromised nodes. This structure provides us tools to analyze per-node-behaviors of distributed algorithms under adversarial environments. The transmissions of misleading information $\widehat{\mathbf{r}}_v$ can be also analyzed via the connections between neighboring games. 
\section{Impact of Attacks}
In this section, we present numerical experiments on DSVM under adversarial environments. We will verify the effects of both the training attacks and the network attacks. {The performance of DSVM is measured by both the local and global classification risks. The local risk at node $v$ at step $t$ is defined as follows:
\begin{equation}
\label{eq:ErrorInNodev}
{{{R}}_v^{(t)}}: = \frac{1}{{{ N_{v}}}}\sum\limits_{n = 1}^{{ N_{v}}} {\frac{1}{2}\left| {{{ y}_{vn}} - {{\widetilde y}_{vn}^{(t)}}} \right|} ,
\end{equation} 
where ${{{y}_{vn}}}$ is the true label, ${{{\widetilde y}_{vn}^{(t)}}}$ is the predicted label, and $N_v$ represents the number of testing samples in node $v$. The global risk is defined as follows:   
\begin{equation}
\label{eq:Error}
{{{R}}_{G}^{(t)}}: = \frac{1}{\sum_{v \in \mathcal{V}} {N_v}}\sum\limits_{v \in \mathcal{V}} {\sum\limits_{n = 1}^{{{ N}_v}} {\frac{1}{2}\left| {{{y}_{vn}} - {{\widetilde y}_{vn}^{(t)}}} \right|} } ,
\end{equation} 
A higher global risk shows that there are more testing samples being misclassified, i.e., a worse performance of DSVM.}

We define the degree of a node $v$ as the actual number of neighboring nodes $|\mathcal{B}_v|$ divided by the most achievable number of neighbors $|\mathcal{V}|-1$. The normalized degree of a node is always larger than 0 and less or equal to 1. A higher degree indicates that the node has more neighbors. We further define the degree of the network as the average degrees of all the nodes. 
\subsection{DSVM Under Training Attacks}
Recall the attacker's constraints $\mathbf{q}_v^T\theta_v \leq Q_v$ in training labels attacks, where $\mathbf{q}_v:=[0,..,0,q_{v(N_v+1)},...,q_{v(2N_v)}]^T$ indicates the cost for flipping labels in node $v$. This constraint indicates that the attacker's modifications in node $v$ are bounded by $Q_v$. Without loss of generality, we assume that $\mathbf{q}_v:=[0,..,0,1,...,1]^T$, and thus, the constraint now indicates that the number of flipped labels in node $v$ are bounded by $Q_v$. We also assume that the attacker has the same $Q_v = Q$ and $C_{v,\delta} = C_\delta$ in every compromised node $v\in\mathcal{V}_a$. Note that we assume that the learner has $C_l=1$ and $\eta = 1$ in every experiments. 

\begin{figure}
\centering
\subfigure[]{
\includegraphics[width=0.45\textwidth]{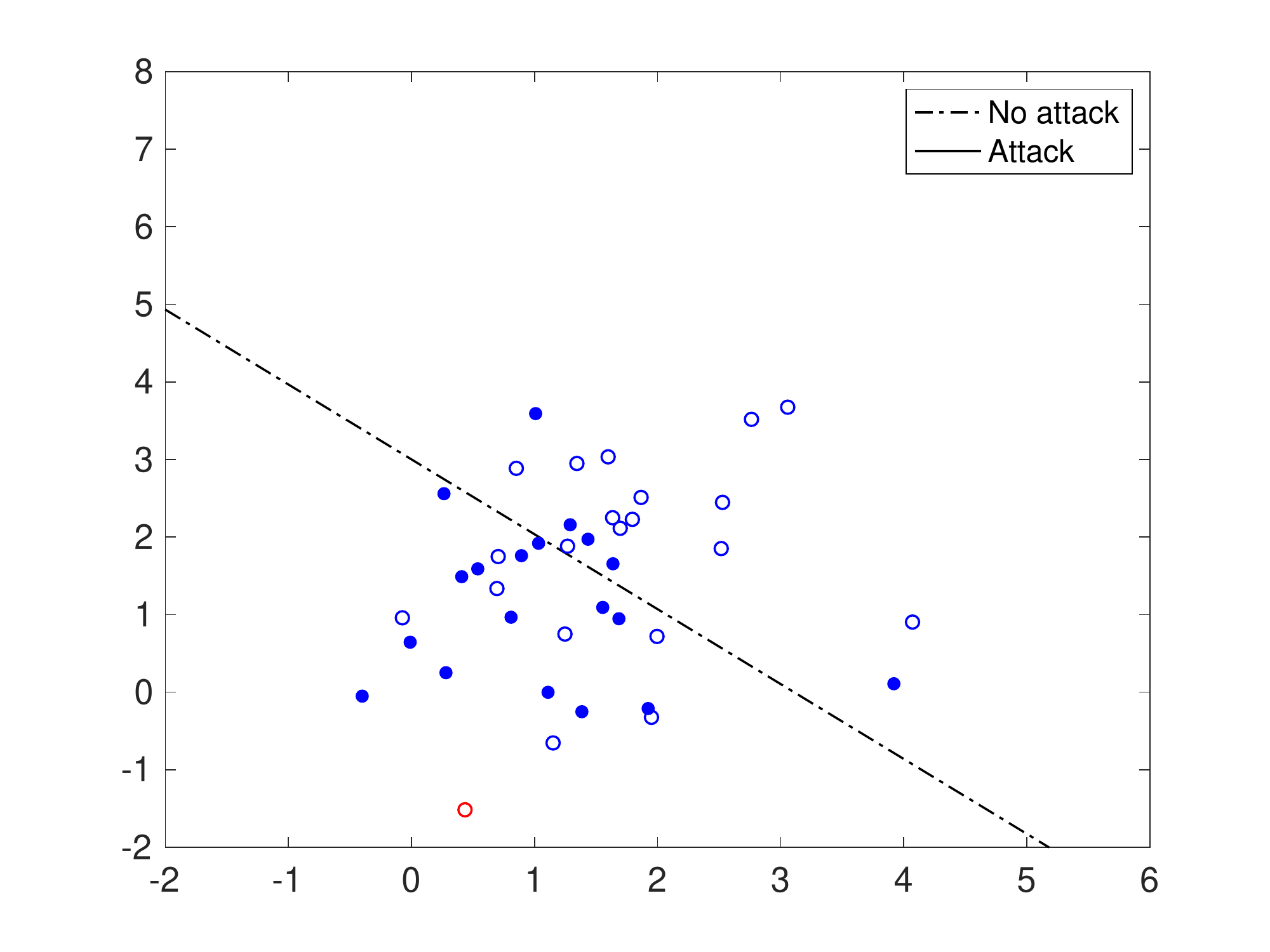}}
\subfigure[]{
\includegraphics[width=0.45\textwidth]{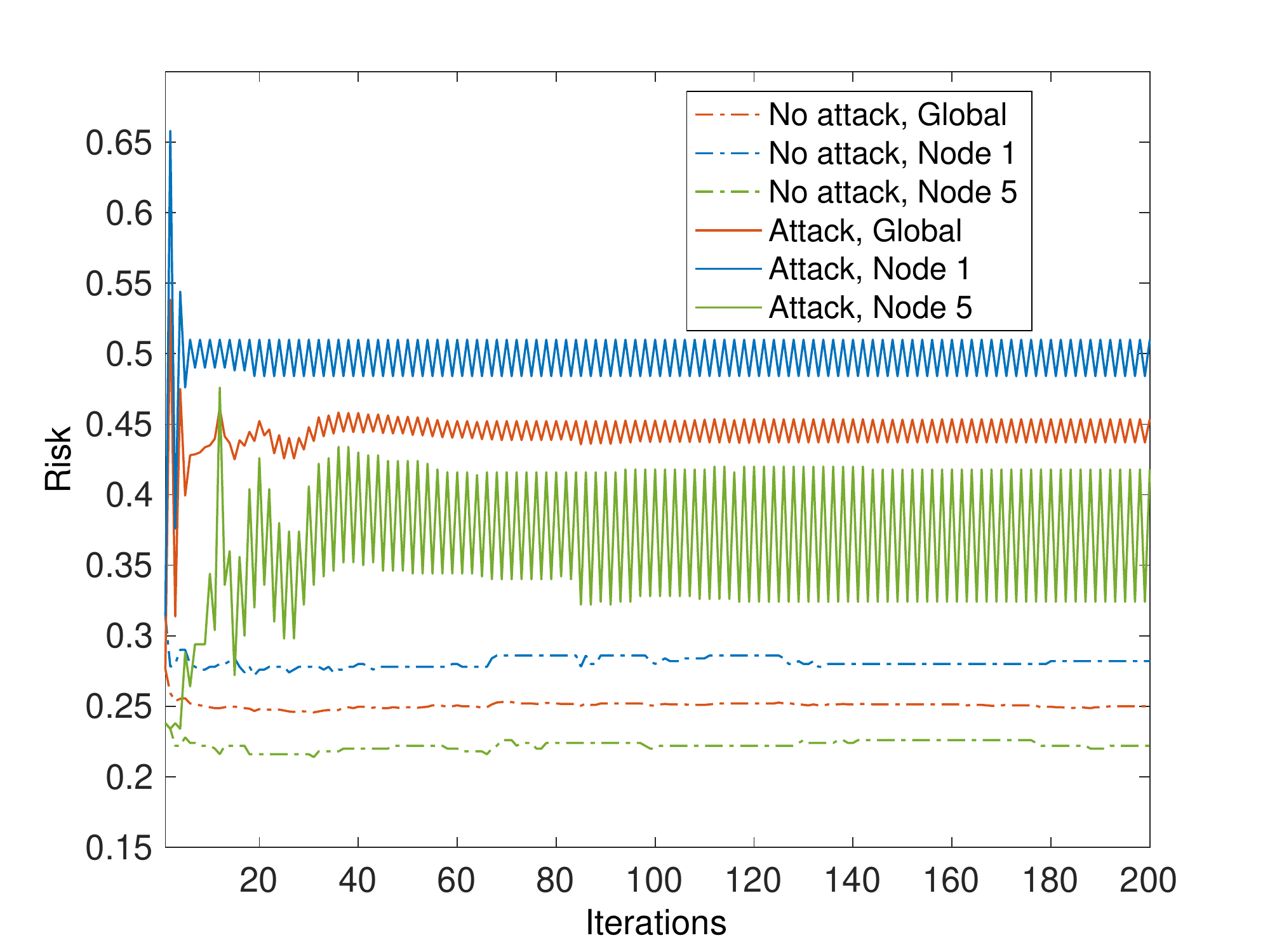}}
\caption{ADMM-DSVM under training-label-attacks with $|\mathcal{V}_a|=4$, $Q=30$, and $C_a=0.01$. }
\label{fig:LabelEx01}
\end{figure}

\begin{figure}
\centering
\subfigure[]{
\includegraphics[width=0.45\textwidth]{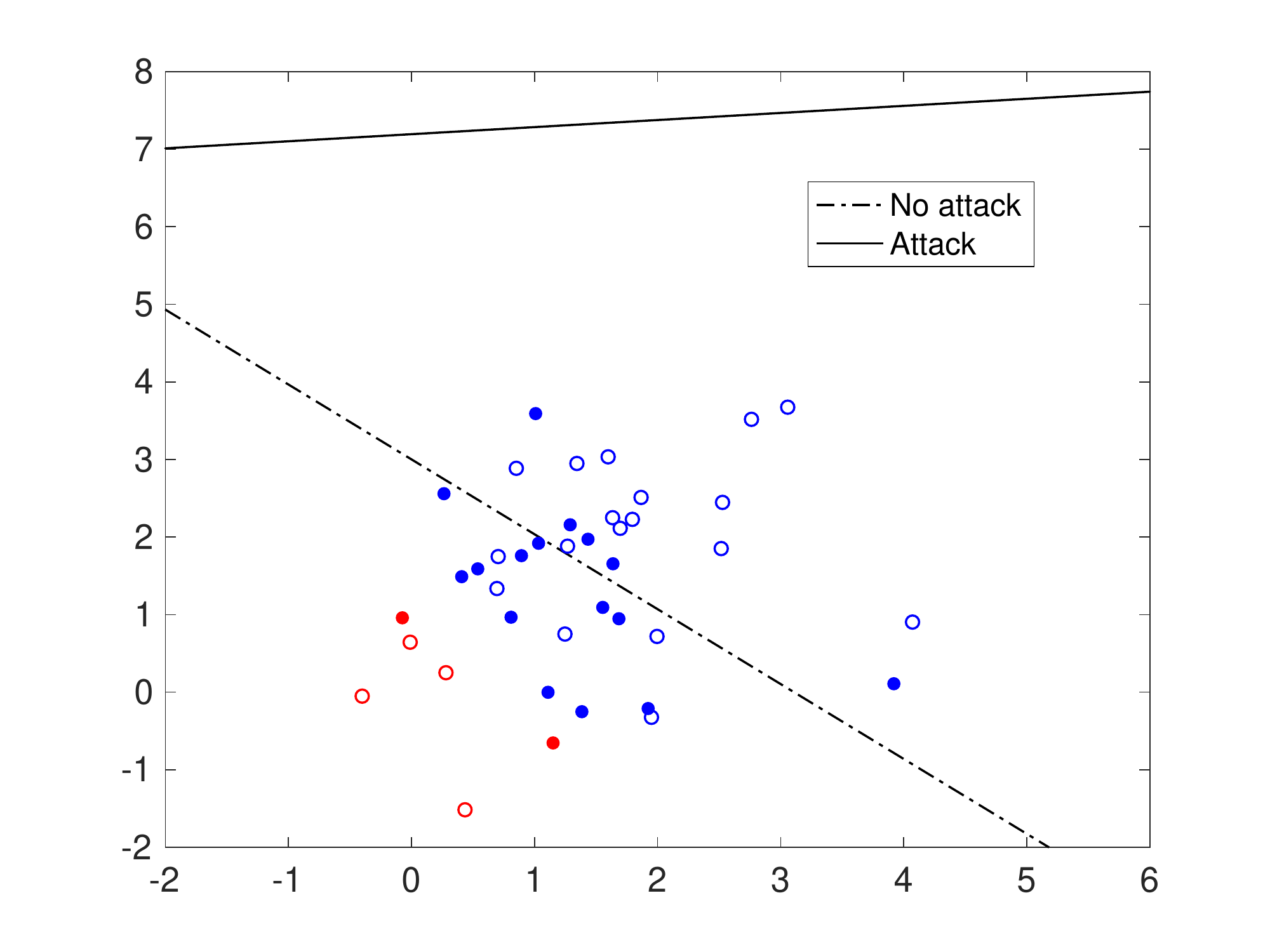}}
\subfigure[]{
\includegraphics[width=0.45\textwidth]{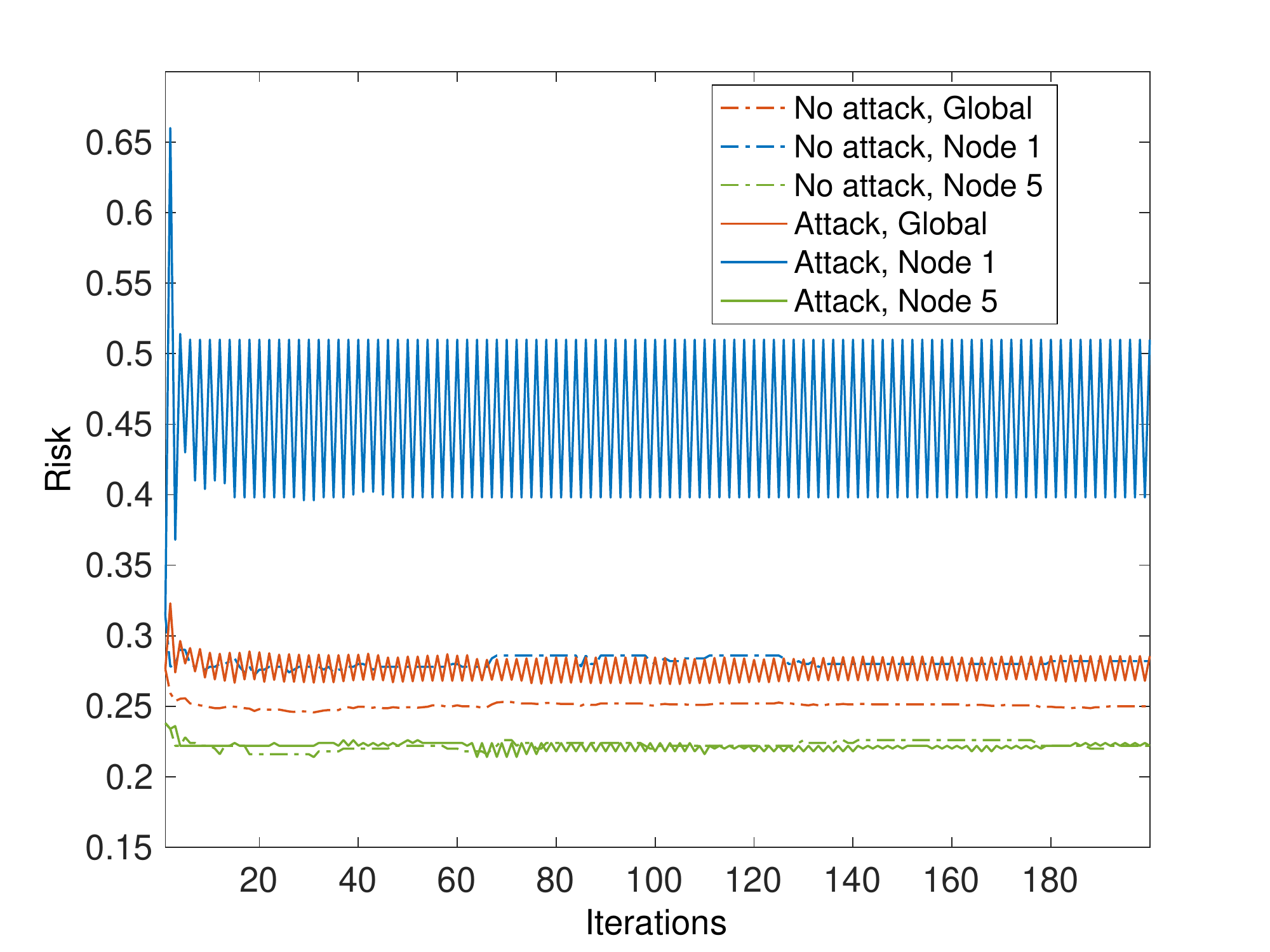}}
\caption{ADMM-DSVM under training-label-attacks with $|\mathcal{V}_a|=1$, $Q=30$, and $C_a=0.01$. }
\label{fig:LabelEx02}
\end{figure}

\begin{figure}
\centering
\subfigure[]{
\includegraphics[width=0.45\textwidth]{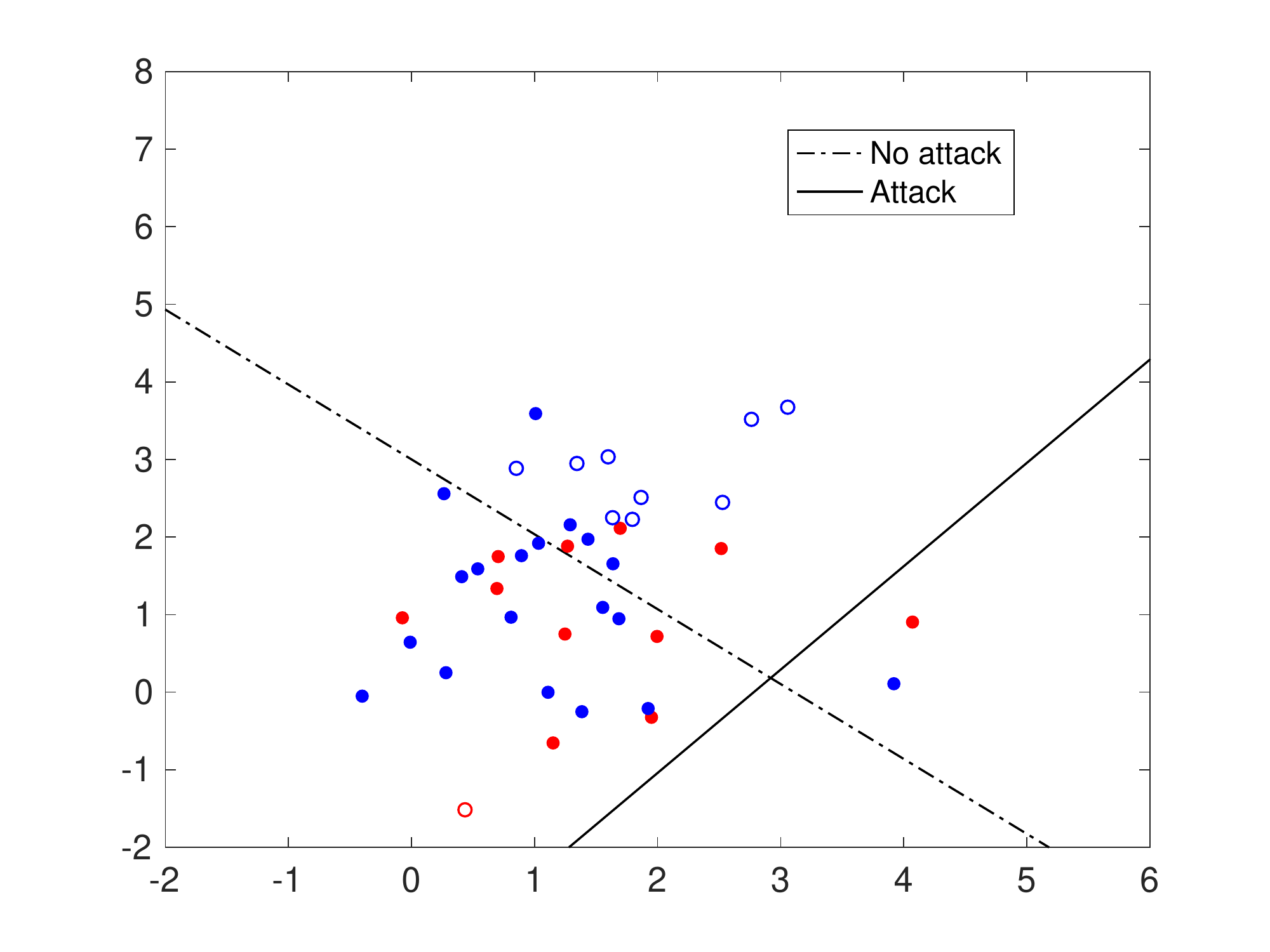}}
\subfigure[]{
\includegraphics[width=0.45\textwidth]{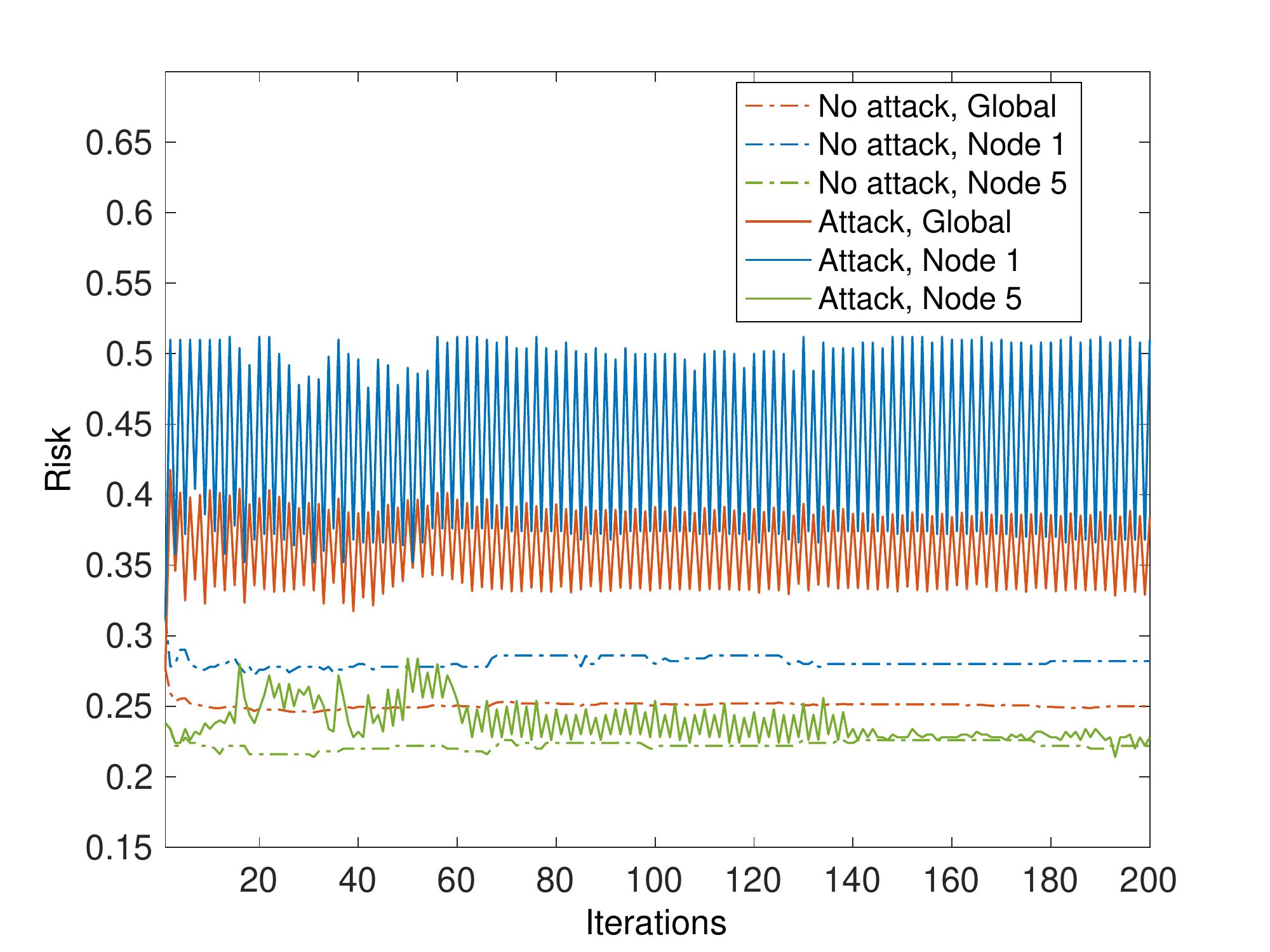}}
\caption{ADMM-DSVM under training-label-attacks with $|\mathcal{V}_a|=4$, $Q=10$, and $C_a=0.01$. }
\label{fig:LabelEx03}
\end{figure}

\begin{figure}
\centering
\subfigure[]{
\includegraphics[width=0.45\textwidth]{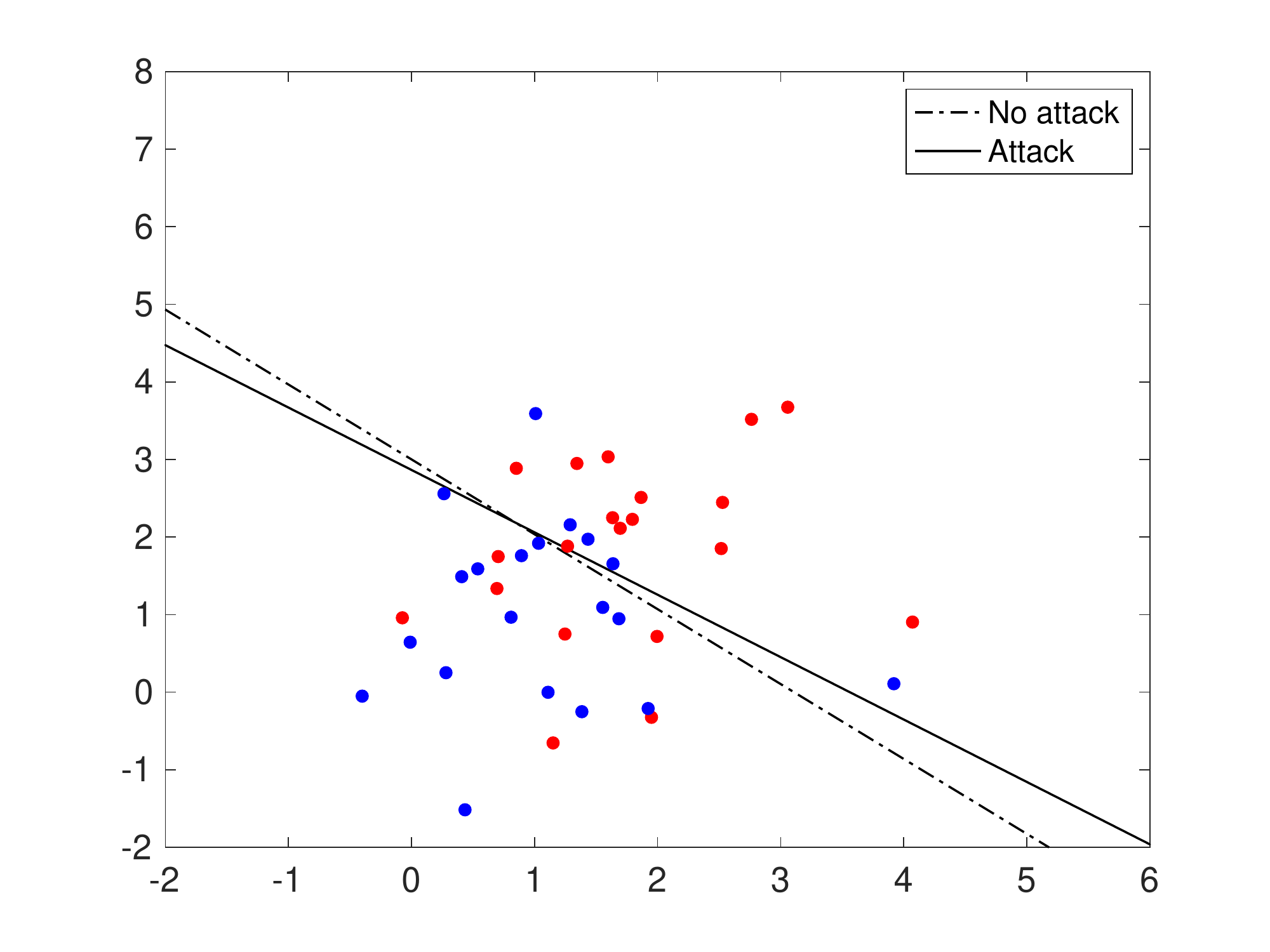}}
\subfigure[]{
\includegraphics[width=0.45\textwidth]{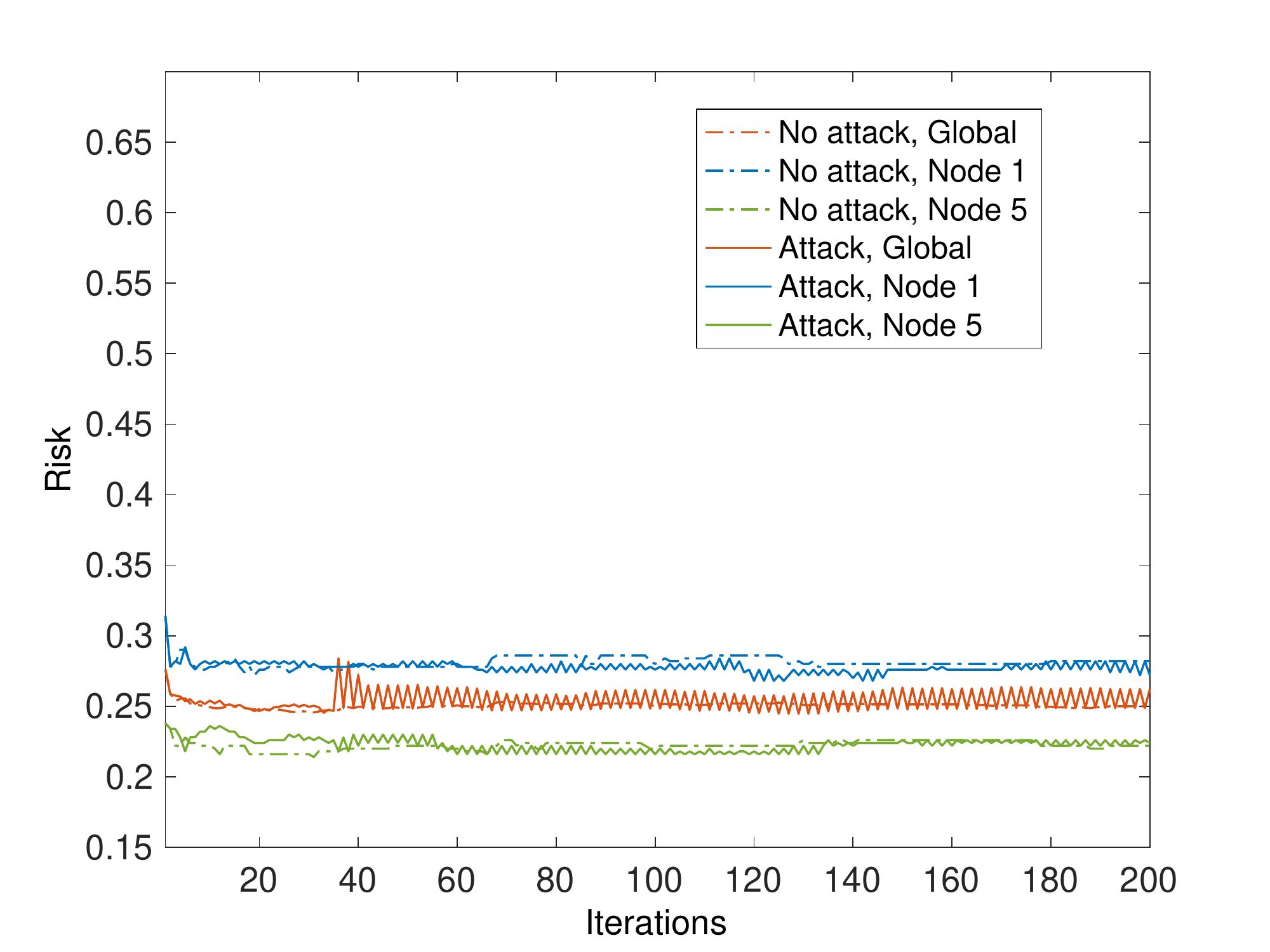}}
\caption{ADMM-DSVM under training-label-attacks with $|\mathcal{V}_a|=4$, $Q=30$, and $C_a=5$. }
\label{fig:LabelEx04}
\end{figure}

Figures \ref{fig:LabelEx01}-\ref{fig:LabelEx04} show the results when the attacker has different capabilities to flip labels in a fully connected network with 6 nodes \cite{zhang2017ciss}. Each node contains $40$ training samples and $500$ testing samples from the same global training dataset with labels $+1$ and $-1$. The data are generated from two-dimensional Gaussian distributions with mean vectors $[1,1]$ and $[2,2]$, and same covariance matrix $[1,0;0,1]$. $|\mathcal{V}_a|$ indicates the number of compromised nodes. $Q$ indicates the largest number of training samples that can be flipped in each node. $C_a$ indicates the cost parameter. Blue filled circles and red filled circles are the original samples with class $+1$ and $-1$, respectively. Blue hollow circles and red hollow circles are the flipped samples with class $+1$ and $-1$, i.e, originally $-1$ and $+1$ respectively. We can see from the figures that when the attacker attacks more nodes, the equilibrium global risk is higher, and the uncompromised nodes have higher risks (i.e., Figures \ref{fig:LabelEx01} and \ref{fig:LabelEx02}). When $Q$ is large, the attacker can flip more labels, and thus the risks are higher (i.e., Figures \ref{fig:LabelEx01} and \ref{fig:LabelEx03}). When $C_a$ is large, the cost for the attacker to take an action is too high, and thus there is no impact on the learner (i.e., Figures \ref{fig:LabelEx01} and \ref{fig:LabelEx04}). Note that, when the attacker has large capabilities, i.e., larger $Q$, smaller $C_a$ or larger $V_a$, even uncompromised nodes, e.g., Node $5$, has higher risks. 

\begin{table}
\caption{Average equilibrium classification risks $(\%)$ of DSVM using Spambase dataset \cite{Spambase} in Network 1 and Network 2.  }
\label{tab:LabelNetwork12}
\begin{center}
\begin{small}
\begin{sc}
\begin{tabular}{|c|c|c|c|c|c|c|}
\hline
Net & 1 & 1L & 1D & 2  & 2L & 2D \\
\hline
Risk & 11.6 & 32.3 & 42.2 & 10.6 & 29.4& 39.3 \\
\hline
STD & 1.6 & 0.6 & 2.6 & 0.6 & 0.3 & 1.1 \\
\hline
\end{tabular}
\end{sc}
\end{small}
\end{center} 
\vskip -0.1in
\end{table}

\begin{table}
\caption{Average equilibrium classification risks $(\%)$ of DSVM using Spambase dataset \cite{Spambase} in Network 3 and Network 4.  }
\label{tab:LabelNetwork34}
\begin{center}
\begin{small}
\begin{sc}
\begin{tabular}{|c|c|c|c|c|c|c|c|c|}
\hline
Net & 3 & 3La & 3Lb & 3Da & 3Db &  4  & 4L & 4D \\
\hline
Risk & 11.7 & 29.5 & 26.9 & 36.4 & 34.6 & 13.5 & 35.0 & 47.0\\
\hline
STD & 1.5 & 0.6 & 1.2 & 0.9 & 0.8 & 1.8 & 0.9 & 2.5\\
\hline
\end{tabular}
\end{sc}
\end{small}
\end{center} 
\vskip -0.1in
\end{table}

Tables \ref{tab:LabelNetwork12} and \ref{tab:LabelNetwork34} show the results when the learner trains on different networks. $L$ and $D$ indicate training labels attacks and training data attacks, respectively. Networks $1$ , $2$ and $3$ have $6$ nodes, where each node contains $40$ training samples. Network $1$ and $2$ are balanced networks with degree $0.4$ and $1$, respectively. The normalized degrees of each node in Network $3$ are $1,0.4,0.4,0.2,0.2,0.2$. Network $4$ has $12$ nodes where each node contains $20$ training samples and has $2$ neighbors. In $1L$ (resp. $1D$) and $2L$ (resp. $2D$), the attacker attacks $6$ nodes with $Q = 20$ (resp. $C_\delta = 10^{10}$). In $3L_A$ (resp. $3D_A$) and $3L_B$ (resp. $3D_B$), the attacker attacks $3$ (resp. $2$) nodes with higher degrees and lower degrees with $Q = 20$ (resp. $C_\delta = 10^{10}$). In $4L$ (resp. $4D$), the attacker attacks $12$ nodes with $Q=10$ (resp. $C_\delta = 0.5\times 10^{10}$). Comparing $1L$ (resp. $1D$) with $2L$ (resp. $2D$), we can see that network with higher degree has a lower risk when there is an attacker. From $3L_A$ (resp. $3D_A$) and $3L_B$ (resp. $3D_B$), we can tell that the risks are higher when nodes with higher degrees are under attack. From $1L$ (resp. $1D$) and $4L$ (resp. $4D$), we can see that network with more nodes has a higher risk when it is under attack. Thus, from Tables \ref{tab:LabelNetwork12} and \ref{tab:LabelNetwork34}, we can see that network topology plays an important role in the security of the ML learner.  
\subsection{DSVM Under Network Attacks}
In this subsection, we use numerical experiments to illustrate the impact of network attacks. In node capture attacks, we assume that the attacker controls a set of nodes $\mathcal{V}_a$, and he has the ability to add noise to the decision variables $\mathbf{r}_v$. In Sybil attacks, the attacker can obtain access to compromised node $v\in\mathcal{V}_a$, then he generates another malicious node to exchange information with the compromised node. Instead of sending random information, which can be easily detected, we assume that he sends a perturbed $\mathbf{r}_v$ to make compromised node believe that this is a valid information, where $\mathbf{r}_v$ comes from the compromised node $v$. In MITM attacks, we assume that the attacker creates adversarial nodes in the middle of a connection, and he receives $\mathbf{r}_v$ from both sides, but he sends a perturbed $\mathbf{r}_v$ to the other sides. 
\begin{figure}[]
\centering
\subfigure[Network 1]{
\includegraphics[width=0.43\textwidth]{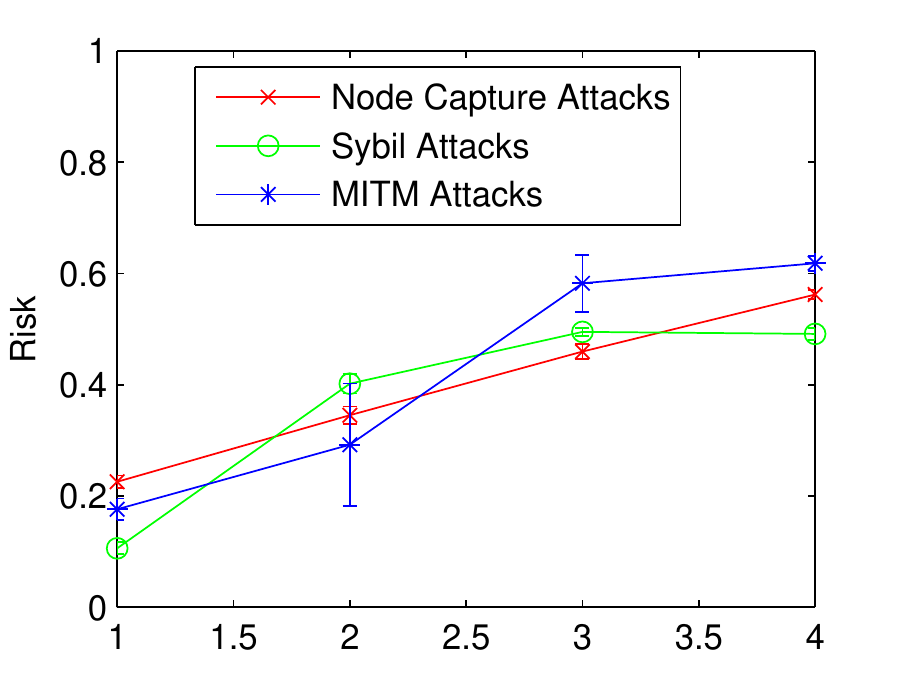}}
\subfigure[Network 2]{
\includegraphics[width=0.43\textwidth]{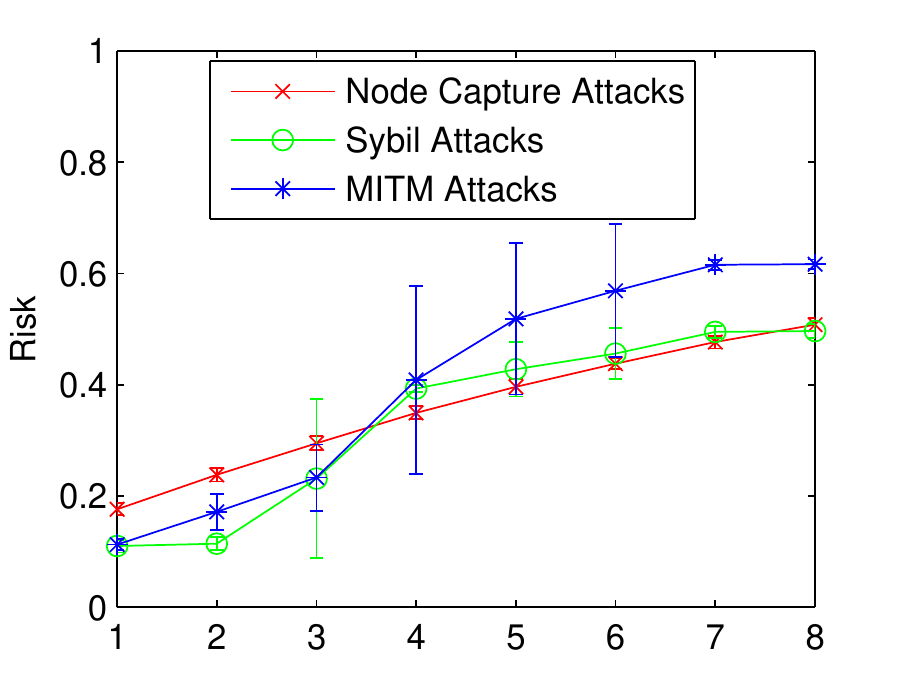}}
\caption{ADMM-DSVM under network attacks.  }
\label{fig:NumericalCyberAttacks}
\end{figure}

{In the experiment in Figure \ref{fig:NumericalCyberAttacks}}, we assume that the elements of the noise vector are generated by a uniform distribution in $[0,R]$, where $R$ indicates the size of the noise. Network 1 and 2 are fully connected networks with $4$ nodes and $8$ nodes, respectively. In node capture attacks and Sybil attacks, the x-axis indicates the number of compromised nodes. The attacker has $R = 1,0.01$ for  node capture attacks and Sybil attacks, respectively. In MITM attacks, the the x-axis represents the number of compromised connections, especially, $1$ indicates only $1$ connection has been broken, $4$ and $8$ for network $1$ and $2$ indicates that all the connections, i.e., $6$ and $28$ connections, have been attacked. Note that the attacker has $R= 0.05,0.02$ for network 1 and 2 in MITM attacks, respectively . From the figure, we can see that when the attacker attacks more nodes or connections, the risk becomes larger. Note that, when more than half of the nodes or connections are compromised, the DSVM will completely fail, i.e, the classifier is the same as the one that randomly labels testing data. 
\section{Discussions and Future Work}
Distributed machine learning (ML) algorithms are ubiquitous but inherently vulnerable to adversaries. This paper has investigated the
security issue of distributed machine learning in adversarial environments. Possible attack models have been analyzed and summarized into machine learning type attacks and network type attacks. Their effects on distributed ML have been studied using numerical experiments.

One major contribution of this work is the investigation of the security threats in distributed ML. We have shown that the consequence of both ML-type attacks and Network-type attacks can exacerbate in distributed ML. We have established a game-theoretic framework to capture the strategic interactions between a DSVM learner and an attacker who can modify training data and training labels. By using the technique of ADMM, we have developed a fully distributed iterative algorithms that each node can respond to his neighbors and the adversary in real-time. The developed framework is broadly applicable to a broad range of distributed ML algorithms and attack models.

Our work concludes that network topology plays a significant role in the security of distributed ML. We have recognized that a balanced network with fewer nodes and a higher degree is less vulnerable to the attacker who controls the whole system. We have shown that nodes with higher degrees can cause more severe damages to the system when under attack. Also, network-type attacks can increase the risks of the whole networked systems even though the attacker
only add small noise to a small subset of nodes.

Our work is a starting point to explore the security of distributed ML. In the future, we aim to extend the game-theoretic framework to capture various attack models and different distributed ML algorithms. We have proposed four defense methods in \cite{zhang2019jaif} to protect DSVM against training data attacks, and we intend to explore other defense strategies for distributed ML by leveraging techniques developed in robust centralized algorithms \cite{globerson2006nightmare,biggio2011support} and detection methods  \cite{levine2006survey,chen2008securing}. {Furthermore, we could use cyber insurance to mitigate the cyber data risks from attackers \cite{zhang2019flipin,zhang2017jsac}.}

%\begin{acknowledgements}
%\end{acknowledgements}


\bibliographystyle{ieeetr}
\bibliography{template.bib}

\end{document}